\documentclass[a4paper]{JHEP3}
\usepackage{epsfig}
\usepackage{graphicx}
\usepackage{amssymb}
\usepackage{latexsym}
\usepackage{braket}
\usepackage{amsmath}

\newcommand{\be}{\begin{equation}}
\newcommand{\ee}{\end{equation}}
\newcommand{\bea}{\begin{eqnarray}}
\newcommand{\eea}{\end{eqnarray}}

\title{POVM generated quantum trajectories without stochastic differential equations.}
\author{Rutvij Bhavsar\\
Department of Mathematics\\
Department of Mathematics, King's College,
London, Strand, London WC 2R 2LS, United Kingdom.\\
Email:\email{rutvij.bhavsar@kcl.ac.uk}\\} 
\author{N.D. Hari Dass\\
Retd.,IMSc,Chennai,INDIA\\
Email: \email{dass@imsc.res.in}\\}

\abstract{
In this paper we examine in depth the issue of quantum trajectories arising out of repeated generalized(POVM) 
Quantum Non-Demolition(QND) measurements on {\it single} copies of unknown states. After a self-contained introduction to various aspects
of quantum measurements, we discuss generalized (POVM) and quantum non-demolition (QND) measurements, as well as quantum trajectories
generated by repeated such measurements. We then discuss an earlier approach by one of us(NDH) given in 2014 based on Gaussian QND measurement 
operators that addressed the asymptotic behaviour of such trajectories. In particular, that analysis showed the impossibility of
determining the unknown state of a single copy from the statistics of such repeated measurements. 

The essence of our present work is the so called martingale and super-martingale properties of certain observables, and the consequent 
martingale convergence theorem which enables to deduce the asymptotic states along such trajectories.
The main result obtained is that asymptotically all trajectories approach either the non-degenerate eigenstates of the system observable, or,
density matrices spanned by the degenerate eigenstates of the observable. The proofs given by us are very transparent. They follow from
straightforward algebra without invoking highly technical aspects from probability theory. 

A unified treatment of both the degenerate and non-degenerate cases is given with the help of projectors of arbitrary dimensionalities. 
In the degenerate case we reproduce the L\"uders prescription. Additionally, the distribution of the trajectories, labelled by the asymptotic 
projectors, is shown to be given exactly by the Born rule. 

Similar conclusions were reached, earlier to us, by Bauer et al on the one hand, and, by Amini et al on the other. A detailed comparison of 
the three approaches is given. A distinctive feature of all three approaches is that no use is made of stochastic differential equations and 
the conclusions follow directly from quantum mechanics. The key to this is staying with the intrinsically discrete 'time' evolution, avoiding 
a continuous time evolution. Alter and Yamomoto were the first to investigate repeated QND measurements on single copies in unknown states.
We make detailed comparisons with their works too.

We end with a brief discussion of i) the robustness of the results against free evolutions of both the system as well as the probe and ii) the
anti-Zeno aspects of the results.
}


\begin{document}
\section{Introduction and motivations.}
Strong motivations for this work, over a long period of time, have been to re-examine some foundational issues in quantum
theory. One of them being  the centrality of the ensembles in quantum theory, in sharp contrast with classical theories in which in principle 
there is no need for ensembles in the description of individual systems. The other being quantum measurements, which too are fundamentally 
different from measurements in classical physics. However,  even in classical physics, ensembles do play very important roles,
as for example, the Gibbs ensembles in statistical mechanics introduced in 1902 for laying the foundations of statistical mechanics. Of course,
the nature of ensembles in quantum mechanics are fundamentally different than those in quantum mechanics.

Ensembles in quantum mechanics arose in response to Born's statistical interpretation of the Schr\"odinger wave 
function \cite{born1926a,born1926b}. It is supremely ironical that Born's first paper \cite{born1926a} was full of grave errors! Throughout
most of the paper he argued for $|\psi(x)|$ to be the probability for finding the quantum particle in position $\mathrm{x}$ when described
by the wavefunction $\psi(x)$. But in a short footnote, he changed his position to claiming $\psi(x)^2$ to be the probability. We know now
that neither of them is correct! The reader is recommended to read Abraham Pais \cite{paisborn1982} for more on this.

It is worth emphasizing that even before Born's work, the probabilistic nature of quantum mechanics had already drawn explicit attention. In 
the seminal paper of Heisenberg \cite{heisenqm1925} that heralded the birth of quantum mechanics, the expression "quantum theoretical
transition probabilities" explicitly appears in the second para after eqn.(16) in the english translation, and the expression
"quantum theoretischen \"Ubergangswahrscheinlichkeiten" in the corresponding place in the German translation. 

Even more puzzling is the fact that Born and Jordan, in their formulation of quantum mechanics \cite{bornjordanqm1925} immediately following 
Heisenberg's breakthrough, state "...for the assumption Heisenberg made that the squares of the absolute values of the elements of a
matrix representing the electrical dipole moment of an atom provide a measure for the transition probabilities"(just before chapter 1).
Again, in the beginning of their chapter 4 "According to Heisenberg, the square of the absolute value $|q(n,m)|^2$ is definitive of the
jump probabilities". In the english translation included in \cite{vdWaerden} this entire chapter has been mysteriously omitted.

In the light of this, it is very surprising that Born in \cite{born1926a} fumbled so much as to what represents probabilities in quantum
mechanics. That Heisenberg is not given some credit for the probability interpretation is equally surprising.

The earliest premonitions of a statistical nature of the coming quantum theory were actually sounded by Albert Einstein in his seminal
paper \cite{einstein1917}(where he introduced the famous A and B coefficients). In his concluding remarks, Einstein stated that the weakness
of the theory(quantum theory) lies in the fact that it leaves the time and direction of elementary processes to 'chance'.

Soon after Born's proposal for the statistical interpretation of quantum mechanics, Pascual Jordan attempted to provide an axiomatic
foundation for quantum mechanics \cite{jordanfoundqm1927}. He based his axioms on the familiar probability calculus. He introduced the notion of {\it Probability Amplitudes} whose absolute squares represented probabilities in quantum mechanics. He postulated that the amplitudes, not 
their absolute squares, obey the same properties of addition(for mutually exclusive cases) and multiplication(for mutually independent cases)
as do probabilities conventionally.

We now describe von Neumann's seminal contributions to quantum measurement theory. The most frequently cited source is his classic book {\it Mathematical Foundations of Quantum Mechanics} \cite{vNmathfound}. But what is not adequately appreciated is that five years before this
work, he published three foundational papers whose annotated english translations are now available \cite{aduncan2025}. In the first two of
this 'trilogy'(as named by Duncan), Von Neumann not only laid the precise mathematical foundations of quantum mechanics in terms of
Hilbert Spaces, but also formulated the notions of states and measurements in quantum mechanics.

von Neumann critiqued Jordan's axiomatization in \cite{JvN1927a} and found it to be unsatisfactory on both conceptual and technical grounds. 
In particular, he found the 'operational status' of probabilities in Jordan's proposals to be not very clearly specified \cite{aduncan2025}.
Neumann proposed drastic modifications to Jordan's proposals. Most foundational among them being: a) replacing Jordan's probability amplitudes
by elements of a Hilbert space as being more fundamental, b) a thorough and mathematically rigorous treatmentof Hilbert spaces, c) equally 
rigorous treatment of the so called eigenvalue problem which he saw as the very essence of quantization, d) introduction and elaboration of 
the powerful notion of {\it Projection Operators}, e) replacing Jordan's multiplication rule by a {\it law of superposition of probabibility 
amplitudes}, and, f) a formulation of the Born rule in terms of Projection operators as its most precise and mathematically rigorous form.  
Additionally, Neumann showed the complete identity of wave mechanical and matrix mechanical descriptions of quantum mechanics complementing 
similar demonstrations by Schr\"odinger, Dirac, and Jordan. 

Ensembles were introduced in the precise sense of Von Mises in von Neumann's classic second paper \cite{JvN1927b}, which also gave a very
clear treatment of the quantum measurement problem. In particular, the state reduction hypothesis, which came to be identified as the 
{\it Collapse Postulate}, and a key element of the so called Copenhagen interpretation, makes its appearance for the first time in this paper.
von Neumann uses ensembles in this precise sense to give a frquentist meaning to probabilities in quantum mechanics.  

After introducing ensembles, von Neumann seeks ways of gaining knowledge about them. Fiven a system S, an ensemble is a collection $S_1,S_2,\ldots$ of the system in various possible {\it States}. He introduces the notion of states somewhat indirectly at first but refines their meaning
considerably in the course of the power. For any observable ${\bf a}$ of the system, measurement of ${\bf a}$ on $S_1$ would yield the value $a_1$ etc. thus the ensemble generating an observed {\it distribution} of values. At this stage, even the notion of measurements and outcomes
is not very precisely described. Consequently the observed {\it expectation value} $\mathbb{E}[{\bf a}]$ can be obtained from the observed
distribution of values in much the same way as in usual probability calculus. von Neumann asserts that full 'knowledge' about an ensemble is in the collection of all possible expectation values $\mathbb{E}[{\bf a}]$ for every observable of the system.

He then goes on to stating various properties of expectation values based on classical probability theory as axioms. Most remarkably, he
develops a statstical operator which encodes all information about an ensemble in the sense that it determines the expectation values of
any observable. This is what has now come to be known as the {\it Density Matrix}.

He also introduced a variety of ensembles. At one end are the so called {\it Elementary Random Ensembles} in which every state of the system
is equiprobable. These are the maximally mixed states and von Neumann shows that their density matrix is a constant multiple of the unit
operator. He also characterises such ensembles as the ones in which one knows nothing about the system i.e. total ignorance. 

At the other end, he introduced the so called {\it Uniform} or {\it Pure} ensembles where every element of the ensemble is in the same state. 
He established the crucial result that the density matrix for the pure state represented by the vector $\phi$ in the Hilbert space is the
projection operator $P_\phi$ and consequently the expectation value of the observable ${\bf a}$ of the system represented by the operator
A on the Hilbert space is given by $\mathbb{E}[{\bf a}]\,=\,Tr P_\phi\,A\,=\,\langle\,\phi|A|\phi\rangle$, in Dirac's notation. This also
established the crucial aspect that states in quantum mechanics are not represented by vectors in Hilbert space and they are actually
represented by the corresponding projection operators, also called the projective or ray representation.

Between these two extremes, von Neumann shows how various ensembles can be arrived at by suitably mixing subensembles of the elementary
random ensemble. These form the totality of all states of the quantum theory including both pure and mixed states.
 
This paper\cite{JvN1927b} is full of many foundational results and it is beyond the scope of this paper to discuss all of them. We shall
focus on those that have a direct bearing on this paper.

Among the important ensembles(states) that he considers are those that have the same outcomes for every measurement on them. In other words,
the probability distribution becomes sharp and the expectation value coincides with the particular value obtained in every measurement. He
then proves another crucial result (stated as point $\delta$ in sec.(3.4) that the necessary and sufficient condition that the statistical
distribution of quantity ${\bf a}$ with associated operator A be sharp is that the vector $\phi$ of the (pure) ensemble  be an eigenstate
of A; the sharp value then is the corresponding eigenvalue.

Equally important, was von Neumann's recognition that there are operators of which $\phi$ is not an eigenfunction, so there must, for any
state, exist quantities for which the statistical distribution is sharp. This has the implication that there are elements of the ensemble
for which the outcomes are different.

He goes on to address another very critical issue, namely, the state of a system subsequent to a measurement. At this point, he introduces
the notion of {\it Repeatability} of measurements even in quantum theory. In his words, an immediate repetition of the experiment will give
the same results. He elaborates the operational meaning of this repeatability criterion  in his footnote 30(on p.99 of \cite{aduncan2025}) 
as well as in the summary in sec.9(on p.107 of \cite{aduncan2025}).
Classically, such a criterion is invoked to test the reliability or consistency of the measurements.

He does not elaborate what exactly he means by 'the same results'. There are in principle two distinct possibilities here: one is that on any 
given element of the ensemble, repeated measurements yields identical values. The other is that, performed on the entire ensemble, repeated
measurements produce the same statistical distribution of values. However, the first is a sufficient condition for the second, but not a
necessary one.

Interpreting repeatability in the first sense has some dramatic consequences. Repeating the experiment many times on the same element would
then go on yielding the same value, resulting in a sharp statistical distribution. But according to von Neumann's earlier demonstration, this
would only be possible if the state immediately after the first measurement is an eigenstate of the observed quantity. It should be carefully
noted that this does not apply to the first measurement itself.

This in essence is his postulate of state reduction, a cornerstone of modern quantum theory. It also got the unfortunate tag as the
collapse of the wavefunction. It represents a true physical process whereby one state of the system goes over into another. Unlike state 
changes in classical physics, this change is effected through the intervention of a measurement.

As already stated, von Neumann had pointed out that generically there are operators representing physical quantities of which a given
element $\phi$ of the Hilbert space is not an eigenfunction and therefore there must, for any state, exist quantitities whose statistical
distributions are not sharp. The Born rule, more precisely stated and inductively proved by him, would determine the probabilities of
obtaining different eigenvalues and consequently with the corresponding eigenstates to be the states resulting from the state reduction.

This in a nutshell is quantum measurement theory according to von Neumann. It was clear that Born's statistical interpretation had no
operational meaning without a specification of the measurement process itself in quantum theory. Most notable steps in this direction
were Heisenberg's paper on uncertainty \cite{heisenunc1927} and Bohr's famous Como lecture \cite{bohrqpost1928} on the quantum postulate.
Heisenberg's paper, apart from making explicit the value-eigenvalue connection, also states quite clearly the notion of state reduction too.
Bohr's paper on the other hand is not very explicit on these issues. Neither Heisenberg's uncertainty paper nor Bohr's quantum postulate
paper say anything about the repeatability assumption.  von Neumann's second paper \cite{JvN1927b} stands out for a mathematically clear and 
precise formulations of these two conceptual pillars for the first time.

Interestingly, von Neumann, in his book \cite{vNmathfound}, cites an earlier experiment by Compton and Simons \cite{comptonsimons1925}
in support of his state reduction hypothesis. But his \cite{JvN1927b} makes no mention of this experiment. Later on, Braginski and Khalili
\cite{brakha1992} have also highlighted this experiment in this context. Recently, R.N. Sen has challenged 
this interpretaion of von Neumann \cite{rnsen2022}. 

This particular type of measurements came to be known as {\it Projective Measurements}. The key assumption of von Neumann leading to
this class of measurements is that of repeatability. This, along with von Neumann's proof that only eigenstates can give rise to sharp 
statistical distribution(of outcomes), led to the state immediately after a measurement being one of the eigenstates. Thus, the act of 
measurement acts as a projection(onto eigenstates). We will soon see that the so called generalized measurements(POVM for Positive Operator
Valued Measurements) do not satisfy the repeatability assumption.

At this stage it is important to recognize two distinct aspects of state reduction in von Neumann's measurement theory. The first is what
happens to each distinct element of the ensemble which we shall henceforth call measurements on a single copy. The other is what happens
to an entire ensemble after a measurement on each of its elements which we shall call ensemble measurements. Focussing on a pure ensemble
each of whose elements is in the same state associated with the vector $|\psi\rangle$ of the Hilbert space, and letting $|a_i\rangle$ be
the Hilbert space vectors associated with the eigenstates of the operator $A$ associated with the observable ${\bf a}$, the state reduction
on single copies is
\be
\label{eq:stateredsingle}
|\psi\rangle\,\rightarrow\,|a_i\rangle
\ee
This happens randomly(unpredictably) with probabilities given by
\be
\label{eq:probstatered}
P(a_i)\,=\,Tr\,\Pi_\psi\,\Pi_i
\ee
where $\Pi_\psi,\Pi_i$ are the projection operators for the state vectors $|\psi\rangle,|a_i\rangle$, respectively. As per von Neumann,
these are also the density matrices(statistical operators) $\rho$ for the corresponding quantum states. Explicitly, 
$\rho_\psi\,=\,\Pi_\psi\,=\,|\psi\rangle\langle\,\psi|$ and likewise, $\rho_i\,=\,\Pi_i\,=\,|a_i\rangle\langle\,a_i|$. It is better to 
recast eqn.(\ref{eq:stateredsingle}) as
\be
\label{eq:redsingle}
\rho_\psi\,\rightarrow\,\rho_i
\ee
and eqn.(\ref{eq:probstatered}) as
\be
\label{probred}
P(\rho_i)\,=\,Tr\,\rho_\psi\,\Pi_i
\ee
The ensemble after the measurement can be determined by averaging the results for single copy measurements over the probability distribution
given. The result is easy to calculate. For that purpose, let us trivially rewrite eqn.(\ref{eq:redsingle}) as
\be
\label{eq:redsingle2}
\rho_\psi\,\rightarrow\,\rho_i\,=\,|a_i\rangle\langle\,a_i|\,=\frac{\Pi_i\rho_\psi\Pi_i}{Tr \rho_\psi\,\Pi_i}
\ee
According to von Neumann, the post-measurement ensemble is obtained by mixing the ensembles corresponding to the eigenstates weighted by
the respective probabilities to get:
\be
\label{eq:postmeasensemble}
\rho_f\,=\,\sum_i\,\frac{\Pi_i\,\rho_\psi\,\Pi_i}{Tr \rho_\psi\,\Pi_i}\cdot Tr \rho_\psi\Pi_i\,=\,\sum_i\,\Pi_i\rho_\psi\Pi_i
\ee
While von Neumann's treatment is fine when the observable has non-degenerate spectrum such that for each eigenvalue there is a unique
eigenstate(modulo an overall phase), the situation gets more confusing in the case of observables with degenerate spectrum, where for
a given eigenvalue there are higher dimensional subspaces, not just one eigenstate, associated with that eigenvalue. In fact, von Neumann's
prescription for state reduction in this case turns out to be incorrect.

L\"uders instead proposed the generalisations \cite{luders1951}:
\be
\label{eq:redsingleluders}
\rho_\psi\,\rightarrow\,\frac{\Pi_{\bar{i}}\rho_\psi\Pi_{\bar{i}}}{Tr \Pi_{\bar{i}}\rho_\psi\,\Pi_{\bar{i}}}
\ee
for state reduction for single copies with probabilities $Tr \Pi_{\bar{i}}\rho_\psi\Pi_{\bar{i}}$,and,
\be
\label{eq:ensembleluders}
\rho_f\,=\,\sum_{\bar{i}}\,\frac{\Pi_{\bar{i}}\,\rho_\psi\,\Pi_{\bar{i}}}{Tr \rho_\psi\,\Pi_{\bar{i}}}\cdot Tr \Pi_{\bar{i}}\rho_\psi\Pi_{\bar{i}}\,=\,\sum_{\bar{i}}\,\Pi_{\bar{i}}\rho_\psi\Pi_{\bar{i}}
\ee
for ensembles. Here $\Pi_{\bar{i}}$ is the $d_{\bar{i}}$-dimensional projection operator onto the degenerate subspace labelled by ${\bar{i}}$
with complex dimension $d_{\bar{i}}$(much more on this later on). When $d_{\bar{i}}=1$ everything reduces to the non-degenerate cases
treated by von Neumann.

There are important differences between the state reduction rules at the individual(in the sense of a particular element of the ensemble) and
the (entire) ensemble. As can be seen from eqn.(\ref{eq:ensembleluders}), the latter is a linear map in state space. On the other hand,the
former, as seen from eqn.(\ref{eq:redsingleluders}), is non-linear.

Quite clearly, von Neumann's projective measurements are not the most general type of quantum measurements as later developments have
shown. The repeatability assumption is specific to the projective measurements. The so called generalized measurements, also called
POVM, do not satisfy this as the states after measurement are in general not the eigenstates of the observable. The projection operators
of the projective measurements get replaced by {\it measurement operators} which do not satisfy the orthogonality properties as
projection operators do. 
But even for generalized measurements, the state reductions (in the sense of the states after measurements) take on the same forms as shown above for projective measurements, with projection
operators being replaced by measurement operators, as will be discussed extensively in Sec.(\ref{sec:povm}). But the crucial features
of non-linearity for individual measurements, and linearity for ensemble measurements,remain.

It is by now clear as to why ensembles are indispensable for state determinations with projective measurements. On a single copy, the
first measurement will lead to one of the eigenstates randomly, and all subsequent measurements, by the repeatability assumption, keep
yielding the same result as the first measurement.

With so called weak measurements, which are limiting cases of generalized measurements (this is made precise in an exactly solvable von
Neumann model in sec.(\ref{sec:gaussianpovm})), there appears to be a way out, at least in principle. 
In such measurements, the state of the system changes only 'weakly'(most of the time, not always; probabilities of large changes are small, 
by design). The outcomes are no longer the eigenvalues. Nevertheless, they do have information, albeit small, about the initial state. How 
exactly this works out will be elaborated shortly. 

Therefore it is conceivable that a large number of sequential weak measurements, done on a {\it single copy}(again unknown) could determine
the original unknown state by using the statistics of the large number of outcomes. Ensembles of identically prepared copies would no longer 
be a necessity then! This would then be a dramatic shift for quantum theory. Of course, the large number of outcomes arising out of single
measurements on single copies in an ensemble measurements would now be replaced by outcomes of a large number of sequential measurements on
a single copy.

There is a curious consistency in quantum mechanics with regard to ensembles vs single copies. Suppose it were possible to make identical
copies of {\it unknown} quantum states much as a classical xerox machine would produce any number of perfect copies from a single original,
irrespective of what the original is, then one could produce a uniform ensemble of quantum states from a single copy. Then , an ensemble
measurement could be made to determine the unknown original state of a single copy. Amazingly, such an {\it Universal Cloning}(Universal in the
sense of being able to do it on unknown states) is impossible in quantum theory!

Ensembles vs single copies has been an issue of concern dating back to Einstein, who eventually accepted the statistical interpretation
of quantum mechanics but held that it only gave an incomplete description of single systems \cite{borneinsteinletts,thornepreface}. In fact
there is extensive discussion of these ideas which may be clubbed under Einstein's {\it Objective Reality} in the Born-Einstein letters, 
which also include Pauli's brilliant explanations of Einstein's worldview in the letters 112, 115 and 116. Diana Buchwald and Kip Thorne have
also given a nice summary in the section on Quantum Mechanics in their preface to the new edition \cite{thornepreface}.

Einstein, as is well known, had hoped that a complete theory would give precisely defined values for all observables, besides restoring
determinism in physics. John Bell's seminal works laid to rest any such hope. Even if one gave a limited interpretation of what Einstein
meant by 'complete description' to mean the ability to determine the unknown state of a quantum system (most unlikely he would have sympathised
with such a view), the main thrust of this work is that such a determination is impossible for single systems. Rather surprisingly, this turns
out to be true not just for weak measurements, but for POVM of arbitrary strength.

Foundational aspects of single state 
measurements were also pointed out by Hartle \cite{hartlesingle1968}. An excellent summary of many results can be found in the book
{\it Quantum Measurement of a Single System} by Alter and Yamamoto \cite{alterbook2001,alter1,alter2}.

As far as the main concerns of this paper, namely, quantum trajectories generated by repeated measurements on a single copy, there have been 
two distinct approaches: the first one as a {\it discrete stochastic process} wherein the state reduction rules define a map from one
stage of the trajectory to the next. The rules themselves are direct consequence of quantum mechanics and therefore, this is a minimalist
approach to the problem.

The other is based on what are called {\it Stochastic Differential Equations} and the concept of {\it Continuous Measurements}. It involves
an idealization where measurements are so dense in time that they are approximated by continuous measurements. Clearly, this formalism
needs assumptions going beyond pure quantum mechanics. It is not at all clear that measurements can be carried out that take arbitrarily
small times. The concept of a stochastic variable that is smooth enough to be differentiable also needs further assumptions. On top of
these difficulties, the stochastic calculi are different depending on where in an infinitesimal time interval the random event take place. Two
most popular schemes are the Ito and Stratanovich calcululi, and they give different results in general. In principle, there can be
infinitely many such.

The stochastic differential equations (SDE) have been developed for both ensemble and single copies. The basic idea is to introduce
stochasticity in time evolution to mimic the randomness arising out of state reductions. As developed by Gisin \cite{gisinprl1984}, and 
subsequently by Diosi \cite{diosi1988}, and, Gisin and Percival \cite{gisinpercival1992}, individual states are described by a state vector 
obeying a SDE. For the corresponding density matrix, the SDE's turn out to be non-linear reflectying the non-linearity of the L\"uders
postulate \cite{luders1951}. The inevitability of this non-linearity has also been clarified by 
Fr\"ohlich \cite{frohlichqm2023,frohlichjumps2025}. We also recommend \cite{jacobscontmeas2006} for an introduction to continuous 
measurements. There is a vast literature on the SDE approach with a number of interesting results and applications. 

We shall, however, work with the minimalist discrete time approach. In a future publication we shall return to the main issues and conclusions 
discussed here from an SDE point of view. 

\subsection{A first attempt.}
A first attempt at addressing the centrality of ensembles in quantum mechanics  was made by one of the authors(NDH) in 
2014 \cite{weakrepeat2014}. But being unaware of how to analyse 
such sequential measurements along a {\it single} trajectory when such trajectories are randomly generated he chose to make an analysis
that essentially amounted to  sequential measurements on ensembles.  He somehow hoped to extract from this what happens to sequential 
measurements on single copies. 

Sequential measurements on a single copy are coded by the sequence of outcomes $p_1,p_2,\dots$, whose particular values characterise an
individual trajectory. The probability of obtaining this particular trajectory is given by $P(p_1,p_2,\ldots)$. What was done in 
\cite{weakrepeat2014} was to calculate, in an exactly solvable gaussian model, the average of $y_M=\frac{\sum_{i=1}^M\,p_i}{M}$ over the 
entire probability distribution $P(p_1,p_2,..)$. This can
be viewed as the joint operations of averaging over both a given trajectory and averaging that average over all the trajectories. The result
obtained, for very large values of M, on the pure state $|\psi\rangle=\sum_i\,\alpha_i|s_i\rangle$ was
\be
\label{eq:weakrepeatensemble}
{\bar y}_M= \sum_i\,|\alpha_i|^2\,s_i=\langle\,\psi|S|\psi\rangle
\ee
exactly as for an ensemble of generalized measurements. But this says nothing about the averages along a particular trajectory. The 
distribution $P(y_M)$ was then calculated the same way and found to be
\be
\label{eq:repeatdistrib}
P(y_M)=\sqrt{\frac{M}{\pi\Delta_p^2}}\sum_i\,|\alpha_i|^2\,e^{-\frac{(y_M-s_i)^2\,M}{\Delta_p^2}}\rightarrow\,\sum_i|\alpha_i|^2\delta(y_M-s_i)
\ee
where $\rightarrow$ indicates the $M\rightarrow\infty$ limit. 
Here $\Delta_p^2$ is a measure of the strength of the generalized measurements. Larger this parameter, weaker are the measurements. {\bf This result
is exact, holding irrespective of whether the measurements are weak or not.} Surprisingly, the asymptotic i.e. the $M\rightarrow\,\infty$
limit is universal and independent of $\Delta_p^2$. We shall explain the meaning of this universality later.
Before interpreting this, let us see what the analogous result would have been
for a single measurement on an ensemble.

In the case of a single measurement on an ensemble, if the probability of the(single)outcome is $P(p)$, the probability of obtaining 
the sequence $p_1,p_2,\ldots$ among $M$ outcomes would be the {\it factorised} form $P(p_1)P(p_2)\ldots$ and the distribution function would take the
form
\be
\label{eq:ensemdistrib}
P_{ens}(y_M)=\sqrt{\frac{M}{\pi\Delta_p^2}}\,e^-{\frac{(y_M-\langle\,S\rangle)^2\,M}{\Delta_p^2}},\quad\quad \langle\,S\rangle=\langle\,\psi|S|\psi\rangle
\ee
In \cite{weakrepeat2014} a {\it Gaussian QND} measurement was used and eqn.(\ref{eq:ensemdistrib}) is an exact result. But for generalized 
measurements this follows on using the Central Limit Theorem, as will be shown later on. This can be done as long as the first and second 
moments of $P(p)$ exist. On the other hand, $P(p_1,p_2,\ldots)$ is not of a factorizable form, but nevertheless the Central limit Theorem can be 
applied to the summands. But we will only use the gaussian case as it suffices to make the essential point. 

Eqn.(\ref{eq:ensemdistrib}) states that for ensemble of single measurements, ${\bar y}_M$ asymptotically approaches the expectation value of 
the observable in the unknown original state. Repeating with {\it optimal} number of observables, at least for finite-dimensional Hilbert 
spaces, the unknown state can be completely determined. On the other hand, for ensemble measurements of a very large number of sequential
measurements, ${\bar y}_M$ as given by eqn.(\ref{eq:repeatdistrib}) asymptotically approaches a sum of delta-function distributions. There
are as many distributions as the number of distinct eigenvalues of the observable being measured, and, each distribution is weighed by a 
factor $|\alpha_i|^2$ which is the probability of finding the corresponding eigenstate or the subspace of degenerate eigenstates in the
original unknown state $|\psi\rangle$.

This was interpreted in \cite{weakrepeat2014} as follows: since the ensemble results, as already emphasized, can be interpreted as first
averaging over a particular trajectory and then averaging that over all trajectories, eqn.(\ref{eq:repeatdistrib}) can be taken to mean
that asymptotically the state along any given trajectory approaches one of the eigenstates,say,$s_i$, and the probability of finding
trajectories labelled by $s_i$ are just those given by the Born Rule i.e. $|\alpha_i|^2$.

This means, surprisingly, repeated generalized measurements on a single state are exactly of the same type as the projective(strong) 
measurements
and are unable to determine the unknown initial state. A given trajectory asymptotically ends up in one of the eigenstates with no information
on the initial state. The information about the original state coded by $|\alpha_i|^2$ is in the distribution of trajectories.

This analysis is only suggestive and not really a proof of the asymptotic behaviour of trajectories. That forms the main body of this work.
It was the work of Maassen and Kummerer \cite{maassen2006} that opened our eyes to the exciting possibility of analysing the behaviour of 
individual trajectories. They showed that under repeated measurements on a single copy, arbitrary initial states(pure or mixed) actually
become pure states almost always. Two essential ingredients to their proof were i) so called Nielsen Identities \cite{nielsen2001}, and,
ii) martingales and the martingale convergence theorem.

In our work, we make use of only the martingale and submartingale properties of certain {\it bounded} quantities along any trajectory, and
use the consequent martingale convergence theorem to show that asymptotically the state of the system tends to one of the eigenstates of the 
observable being measured. Thus our results go beyond Maassen and Kummerer's with fewer inputs. We also show that the distribution of the 
trajectories is given by the Born rule.

After finishing our work we came across closely related works by Bauer and Bernard \cite{bauershort2011}, Bauer,Benoist, and Bernard 
\cite{bauerlong2013}, as well as by Amini,Rouchon, and Mirrahimi \cite{amini2011} obtaining similar results. We shall make a detailed
comparison of these three closely related approaches. All of them are based on repeated Quantum Non-Demolition(QND) measurements. All three
use only {\it discrete time} evolutions without the use of any {\it Stochastic Differential Equations} characterstic of 
{\it continuous} measurements. In all three approaches results follow directly from quantum mechanics.

Already around 1995 Alter and Yamomoto had claimed the impossibility of determining the unknown state of a single system through repeated
quantum measurements \cite{alter1,alter2}. Though their final claims are similar to ours, their methods are very different. Also, 
the asymptotic convergence to arbitrary eigenstates of the system along with their probabilities is not very transparent. A comparison will
 also be made with their approach. Their book
\cite{alterbook2001} contains detailed expositions of these and other related works. 

\section{Generalized measurements(POVM)}
\label{sec:povm}
In this section we introduce the so called {\it Generalized Measurements} also called {\it POVM}(Positive Operator
Valued Measurements). Though most of this is textbook material these days, it is included to make the discussion self-contained and
also to establish notations. It should be recalled that in the early days of quantum mechanics, the type of measurements that dominated the 
discourses were what we now recognize as projective measurements. The salient features of such measurements are i) the outcomes are one of
the eigenvalues of the measured observable,and,ii) the state of the system after measurement becomes the corresponding eigenstate. The latter
is also known as the von Neumann {\it state reduction} postulate. von Neumann gave an explicit mathematical model that realises these
features though the exact mechanism of state reduction remained nnderstood \cite{vNmathfound}. In generalized measuements on the other hand 
neither of these holds. 

Actually the formalism of generalized measurements covers both von Neumann's projective measurements also called
{\it Strong Measurements}, as well as Aharonov's weak measurements. It should however be appreciated that the basic 
framework even for these larger class of measurements is essentially the same as von Neumann's model for measurements \cite{vNmathfound}, 
with many of its foundational difficulties continuing to be unresolved. This follows from the 
Stinespring Dilation Theorem \cite{stinespring1955}, of which the earlier Naimark Extension Theorem \cite{naimark1943} is a special case. 
For proofs and further details, see any standard
text on matrix analysis, such as \cite{bhatia1997}. Here we present von Neumann's measurement model which illustrates how {\it any} POVM
can be realised via a projective measurement on an extended Hilbert space.

In essence, there are three distinct phases in all such models of quantum measurements:i) initial state preparation for both the quantum
{\it system} as well as the {\it probe}. In a loose sense the probe can be thought of as the apparatus measuring some observable in the 
state of the system, but it is better to think of it as an intermediary in the total act of the measurement. What should really be thought
of as an apparatus, in the sense that Bohr envisioned what apparatuses are, will be commented upon later. It is important to keep in mind
that not only the system but the probe is also treated quantum mechanically, ii) a {\it measurement interaction} phase which brings the
system and probe into interaction described by a {\it Unitary} transformation acting on their joint Hilbert space. Often this is confused
for the whole measurement. This is a purely unitary transformation while the complete measurement is not. At the end of this phase the system 
and probe states are {\it entangled}. That is the reason why this phase in itself does not constitute the act of measurement. In order for a 
definite outcome it is essential that the probe and system states are disentangled. Finally, iii) the phase where the system and probe get
disentangled with a definite measurement outcome. Exactly how this last phase comes about is still an outstanding unresolved problem of 
quantum theory,
and forms the crux of the so called {\it Quantum Measurement Problem}.
 
The initial state of the system is taken as general as possible i.e. it can be either a pure state or a mixed state. To accommadate such
generality one adopts the density matrix formulation of quantum states. The initial state of the probe is taken to be pure, for convenience.
The pre-measurement state is taken to be disentangled product of an unknown system state $\theta_0$, and a {probe} state $|\phi_0\rangle_P\langle\,\phi_0|$:
\be
\label{eq:initial}
\rho_0 = \theta_0\otimes|\phi(0)\rangle_P\langle\,\phi_0|
\ee


After the measurement interaction denoted generically by $U$, the joint state of the system and probe is given by
\be
\label{eq:postmeasrho}
\rho^\prime = U\rho_0\,U^\dag
\ee
Let us now introduce the so called {\it Pointer State Basis} which is taken to be an orthonormal basis for the probe Hilbert space 
${\cal H}_P$. These
can be taken to be the eigenstates $|\alpha_I\rangle$ of some operator ${\cal A}$ acting on the probe states. For convenience it will be taken 
to be non-degenerate. As long as ${\cal A}$ remains fixed, and ${\cal H}_P$ is finite-dimensional, these pointer states can also be taken to be
labelled by integers I. It should be noted that Bauer et al use the terminology of pointer states for the states of the system, and not of the
probe as done here. This should be kept in mind while comparing our respective works. Another point to be kept in mind is that in general
${\cal A}$ need not have any connection with the probe observable ${\hat Q}$ appearing in the measurement interaction(it will in fact be seen 
shortly that they actually should not!In von Neumann model they are actually canonically conjugate to each other).

When the choice of ${\cal A}_P$ is changed to another choice ${\cal A}_P^\prime$, the basis changes to $|\alpha^\prime_{I^\prime}\rangle$
which can be equally described as a change in the indexing set $I\rightarrow\,I^\prime$. The eigenstates can also be written as $|I\rangle_P$
etc as long as there is no confusion that I's are necessarily eigenvalues of  ${\hat Q}$.

On using the completeness relation
\be
\label{eq:probecomplete}
\sum_I\,|\alpha_I\rangle_P\langle\,\alpha_I|\,=\,{\bf I}_P
\ee
where ${\bf I}_P$ is the identity on the probe Hilbert space ${\cal H}_P$,
the post measurement-interaction state $\rho^\prime$ can be re-expressed as
\be
\label{eq:postmeas2}
\rho^\prime = \sum_{I,I^\prime}\,|\alpha_I\rangle\langle\,\alpha_I|U\,\theta_0|\phi_0\rangle\langle\,\phi_0|U^\dag\,|\alpha_{I^\prime}\rangle\langle\,\alpha_{I^\prime}|
\ee
Or, equivalently,
\be
\label{eq:postmeas3}
\rho^\prime = \sum_{I,I^\prime}\,\langle\,\alpha_I|U|\phi_0\rangle\,\theta_0\,\langle\,\phi_0|U^\dag\,|\alpha_{I^\prime}\rangle\,|\alpha_I\rangle\,\langle\,\alpha_{I^\prime}|
\ee
Let us now introduce the so called {\it Measurement Operators}, acting on the system states, by
\be
\label{eq:measops}
M_{\alpha_I}=\langle\,\alpha_I|U|\phi_0\rangle_P 
\ee
They satisfy the very important condition
\be
\label{eq:measopunity}
\sum_I\,M_{\alpha_I}^\dag\,M_{\alpha_I}\,=\,{\bf{I}}_S
\ee
For projective measurements, the measurement operators are just projectors onto eigenstates of the observable. Eqn.(\ref{eq:measopunity})
is the analog of {\it Decomposition of Unity} for generalized measurements.
In terms of these $\rho^\prime$ can be expressed as 
\be
\label{eq:postmeasop}
\rho^\prime=\sum_{I,I^\prime}\,M_{\alpha_I}\theta_0\,M_{\alpha_{I^\prime}}^\dag\,|\alpha_I\rangle\langle\,\alpha_{I^\prime}|
\ee
At this stage the system and probe are in general {\it entangled} and the process of measurement is not complete as it does not make sense
to talk of the state of the probe without reference to the state of the system. In fact, the measurement gets completed when the system and
probe are disentangled again so the probe can be measured independent of the system. But it is important to emphasize that the nature of 
disentanglement now
is very different from that of the initial state we started with. Now it is {\it mixed state disentanglement} also called a separable system
and probe state.

How exactly this comes out is still an unresolved issue in quantum mechanics and is the quantum measurement problem. 
There are two possible ways of thinking about how the measurement gets completed.The first of these is {\it Decoherence}, and the second
as a direct or projective measurement of the probe. While both of them lead to identical final results, the direct measurement of the probe
is a sort of an ad hoc prescription as the system and probe are still entangled. We describe both of them now.

\subsection{Decoherence}
The first of these broadly goes by the name {\it Decoherence} and the underlying physics idea has been talked about for a very long time.
See, for example, the classic, {\it Quantum Theory} by David Bohm \cite{bohmqt1989}. The idea is that the interaction betwen the probe
and something like a macroscopic measuring device(apparatus,environment) is so complex that it essentially randomizes all the relative phases 
in the
superposition. Effectively this would {\it diagonalize} the post measurement density matrix. See \cite{paroanu2021} to get a feel for how 
decoherence is treated in realistic systems.

In other words, decoherence would result in $\rho^\prime\rightarrow\,\rho_{decoh}$ where $\rho_{decoh}$ is {diagonal} in $I,I^\prime$ i.e. 
the {Pointer State Basis}. The pointer states had been introduced as some orthonormal basis in the probe ${\cal H}_P$ but now decoherence
points to a preferred choice of pointer states. It is quite obvious that the density matrix can not be diagonal in more than one basis unless
they are unitarily equivalent. Therefore decoherence picks a preferred basis modulo unitary equivalence. So it is natural to identify the 
pointer states with this preferred basis.

This naturally raises the question as to what precisely determines the preferred basis during decoherence. Presumably it has to do with
details of the probe-environment(apparatus) details. One can heuristically understand an apparatus in quantum measurements as that
decohering environment which diagonalizes the $\rho^\prime$ for a given measurement interaction $U({\hat q},{\hat Q})$.

Thus, effectively, the post-decoherence composite state can be taken as
\be
\label{eq:rhodecoh}
\rho_{decoh}=\sum_I\,M_{\alpha_I}\theta_0\,M_{\alpha_I}^\dag\,|\alpha_I\rangle\langle\,\alpha_I|
\ee
which amounts to just retaing the diagonal part of eqn.(\ref{eq:postmeasop}). When ${\cal H}_P$ is infinite-dimensional there are, not
surprisingly, many technical complications to this.
\subsection{Projective measurement of the probe}
Because of many unresolved details with the decoherence mechanism, many would like to picture the composite state after measurement is
complete to be simply the result of a {Projective Measurement} or {Direct Measurement} on the probe, without going into the detailed mechanisms for it.
This is in fact how von Neumann chose to describe them in the first place.

Introducing the projection operators $\Pi_I=|\alpha_I\rangle\langle\,\alpha_I|$, the state after the direct measurement is given, according
to this way of looking, by
\be
\label{eq:rhofproj}
\rho_f = \sum_I\,\Pi_I\,U\,\theta_0\,|\phi_0\rangle\langle\,\phi_0|U^\dag\,\Pi_I
\ee
This step can also be viewed as a {Projective} measurement on the probe, also called a {direct} measurement by some.
It is easy to see that $\rho_{decoh}=\rho_f$.
\subsection{The post-measurement state}
Before going further, it is necessary to further rewrite them to bring out their full meaning. This is necessitated by the fact
that $M_{\alpha_I}\theta_0\,M_{\alpha_I}^\dag$ is not properly normalized as a density matrix.
We Introduce the normalized density matrix
\be
\label{eq:thetanorm}
\theta^\prime_I\,=\,\frac{M_{\alpha_I}\theta_0\,M_{\alpha_I}^\dag}{Tr\,M_{\alpha_I}\theta_0\,M_{\alpha_I}^\dag}
\ee
and the normalized probability distributions
\be
\label{eq:proboutcome}
 p(\alpha_I)\,=\,Tr\, M_{\alpha_I}\theta_0\,M_{\alpha_I}^\dag 
\ee
That the probabilities $p(\alpha_I)$ so defined are indeed correctly normalised can be shown on using eqn.(\ref{eq:measopunity}) as
follows:
\be
\label{eq:proboutnorm}
\sum_I\,p(\alpha_I)\,=\,\sum_I\,Tr\,M_{\alpha_I}\,\theta_0\,M_{\alpha_I}^\dag\,=\,Tr\,\sum_I\,M_{\alpha_I}^\dag\,M_{\alpha_I}\theta_0= {\bf{I}}
\ee
The {post-measurement} state can now be written as
\be
\label{eq:postnormalized}
\rho_f = \sum_I\,p(\alpha_I)\,\theta_{I}^\prime\otimes\,|\alpha_I\rangle\langle\,\alpha_I|
\ee

What eqn.(\ref{eq:postnormalized}) tells is that the post-measurement state is first of all a mixed state. It is also a disentangled state 
though
very different in nature from the disentangled initial state of eqn.(\ref{eq:initial}). Such mixed, disentangled states are also called
{\it separable}. In short, the combined state of the system and probe is a separable mixed state.

It can also be viewed as a {\it classical mixture} much like an urn containing balls of different colors. The probabilities $p(\alpha_I)$
are like probabilities in a classical mixture.

It should be noted that the elements of this mixture, $\theta_I\otimes\,|\alpha_I\rangle\langle\,\alpha_I|$ are perfectly correlated. This
is akin to the eigenvalue-eigenstate correlations in projective measurements though the outcome of probe measurements $\alpha_I$(more
specifically when the probe state is $|\alpha_I\rangle$) is no
longer related to the eigenvalues of the system observable. Nor is $\theta_I$ related to the eigenstates. This is the generic situation.

Stated differently, when the outcome of the probe measurement is $\alpha_I$, the system state is reduced to $\theta_I$. These are the
generalizations of the von Neumann postulates for generalized measurements. It is sometimes erroneously claimed that in generalized
measurements there is no wavefunction collapse(more precisely, von Neumann state reduction). This is incorrect as can be seen from our
exposition.

In summary, the outcome $\alpha_I$ occurs with probability $p(\alpha_I)$ and is accompanied by the probe state being reduced to the
pure state $|\alpha_I$,
and, the system state reduced to $\theta_I$(generically mixed).

In presently accepted accounts of quantum measurements, various possible outcomes are supposed to occur randomly, with no event to event
explanations. Does decoherence have any light to throw on this randomness, one of the central features of quantum theory? Since the 
randomization of the phases is ascribed to the very complex interactions between the probe and the environment, it may be reasonable to
expect that they too play a role in determining the event to event description. The state of the environment then would play the role
of some hidden variables, though not of the deterministic and classical variety considered in current hidden variable theories. This may
remain outside the scope of any realistic attempts. 

The whole process of measurement can be effectively thought of as a {\it mapping} in the state space of the system. In other words, if the 
outcome of 
the probe measurement is $\alpha_I$, the system state undergoes the mapping
\be
\label{eq:measmap}
\theta_0\rightarrow\,\theta^\prime_I\,=\,\frac{M_{\alpha_I}\,\theta_0\,M_{\alpha_I}^\dag}{Tr\,M_{\alpha_I}\theta_0\,M_{\alpha_I}^\dag}
\ee
As already emphasized, in these types of measurements there is very little correlation between the apparatus outcomes $\alpha_I$ and 
any of the system eigenvalues, which we denote by $q_i$!

Some important properties of this map are, i) if $\theta_0$ is a pure state so is $\theta^\prime_I$, and, ii) if $\theta_0$ is a mixed state, 
so will $\theta^\prime_I$ be.
\subsection{Repeated Generalized Measurements}
\label{sub:genrepeat}
In this subsection we shall develop the formalism for repeated generalized measurements on a single copy. The results of this subsection form
the backbone for the rest of this paper. The initial state of the system is taken to be $\theta_0$ and the measurement operators are $M_{\alpha_I}$. We build this up step by step.

After the first step, let the outcome be $\alpha_{i_1}$ and the state be mapped to $\theta_1$ given by
\be
\label{eq:state1}
\theta_1\,=\,\frac{M_{\alpha_{i_1}}\theta_0\,M_{\alpha_{i_1}}^\dag}{Tr M_{\alpha_{i_1}}\theta_0\,M_{\alpha_{i_1}}^\dag}
\ee
The probability of obtaining the first outcome $\alpha_{i_1}$ is given by
\be
\label{eq:prob1}
P(\alpha_{i_1})\,=\,Tr M_{\alpha_{i_1}}\theta_0\,M_{\alpha_{i_1}}^\dag
\ee
At the next step, let $\alpha_{i_2}$ be the outcome and $\theta_2$ the resulting state of the system, given by
\be
\label{eq:state2}
\theta_2\,=\,\frac{M_{\alpha_{i_2}}\theta_1\,M_{\alpha_{i_2}}^\dag}{Tr M_{\alpha_{i_2}}\theta_1\,M_{\alpha_{i_2}}^\dag}
\ee
The probability of obtaining $\alpha_{i_2}$ is now {\it conditional} on the first outcome being $\alpha_{i_1}$, with the conditional probability
being given by
\be
\label{eq:prob2|1}
P(\alpha_{i_2}|\alpha_{i_1})\,=\, Tr M_{\alpha_{i_2}}\theta_1\,M_{\alpha_{i_2}}^\dag
\ee
These expressions suggest a way of rewriting them that will be very useful later on.
\be
\label{eq:state1p}
\theta_1\,=\,\frac{M_{\alpha_{i_1}}\theta_0\,M_{\alpha_{i_1}}^\dag}{P(\alpha_{i_1})}
\ee
and,
\be
\label{eq:state2p}
\theta_2\,=\,\frac{M_{\alpha_{i_2}}\theta_1\,M_{\alpha_{i_2}}^\dag}{P(\alpha_{i_2}|\alpha_{i_1})}
\ee
On using eqn.(\ref{eq:state1p}) in eqn.(\ref{eq:prob2|1}), one gets
\be
\label{eq:condprob2|1p}
P(\alpha_{i_2}|\alpha_{i_1})\,=\,\frac{M_{\alpha_{i_2}}M_{\alpha_{i_1}}\,\theta_0\,M_{\alpha_{i_1}}^\dag\,M_{\alpha_{i_2}}^\dag}{P(\alpha_{i_1})}
\ee
This leads to the very important identification
\be
\label{eq:joint12}
P(\alpha_{i_2}|\alpha_{i_1})P(\alpha_{i_1})\,=\,J(\alpha_{i_1},\alpha_{i_2})\,=\,Tr\,M_{\alpha_{i_2}}M_{\alpha_{i_1}}\,\theta_0\,M_{\alpha_{i_1}}^\dag\,M_{\alpha_{i_2}}^\dag
\ee
where we have introduced the {\it Joint Probability Distribution} $J(\alpha_{i_1},\alpha_{i_2})$ as given by the Bayes Theorem. Finally, on
using eqn.(\ref{eq:state1p}) in eqn.(\ref{eq:state2p}), and using the joint probability distribution,  we arrive at yet another very
useful representation for $\theta_2$:
\be
\label{eq:state2pp}
\theta_2\,=\,\frac{M_{\alpha_{i_2}}M_{\alpha_{i_1}}\,\theta_0\,M_{\alpha_{i_1}}^\dag\,M_{\alpha_{i_2}}^\dag}{J(\alpha_{i_1},\alpha_{i_2})}
\ee
We now seek the generalizations of these to arbitrary number of repeated mesurements. The generalization of eqn.(\ref{eq:state2p}) is
straightforward:
\be
\label{eq:staten}
\theta_n\,=\,\frac{M_{\alpha_{i_n}}\theta_{n-1}\,M_{\alpha_{i_n}}^\dag}{P(\alpha_{i_n}|\alpha_{i_1}\ldots\alpha_{i_{n-1}})}
\ee
with the conditional probability of $\alpha_{i_n}$ given $(\alpha_{i_1}\ldots\alpha_{i_{n-1}})$ being given by
\be
\label{eq:ncondprob}
P(\alpha_{i_n}|\alpha_{i_1}\ldots\alpha_{i_{n-1}})\,=\,Tr\,M_{\alpha_{i_n}}\theta_{n-1}\,M_{\alpha_{i_n}}^\dag
\ee
The generalization of eqn.(\ref{eq:state2pp}) and eqn.(\ref{eq:joint12}) are however not that straightforward. Here we prove them by 
induction. Let the following be true for some N:
\be
\label{eq:stateNpp}
\theta_N\,=\,\frac{M_{\alpha_{i_N}}\ldots\,M_{\alpha_{i_1}}\,\theta_0\,M_{\alpha_1}^\dag\ldots\,M_{\alpha_{i_N}}^\dag}{J(\alpha_{i_1},\ldots,\alpha_{i_N})}
\ee
along with
\be
\label{eq:jointN}
J(\alpha_{i_1},\ldots\alpha_{i_N})\,=\,Tr M_{\alpha_{i_N}}\ldots\,M_{\alpha_{i_1}}\theta_0\,M_{\alpha_{i_1}}^\dag\ldots\,M_{\alpha_{i_N}}^\dag
\ee
be the joint probability distribution for the first $N$ outcomes. 

We shall now prove that these eqns also hold for $N+1$. The proof goes as follows:
\bea
\label{eq:stateind}
\theta_{N+1}\,&=&\,\frac{M_{\alpha_{i_{N+1}}}\theta_N\,M_{\alpha_{i_{N+1}}}^\dag}{P(\alpha_{i_{N+1}}|\alpha_{i_1}\ldots\alpha_{i_N})}\nonumber\\
              &=&\,\frac{M_{\alpha_{i_{N+1}}}\ldots\,M_{\alpha_{i_1}}\theta_0\,M_{\alpha_{i_1}}^\dag\ldots\,M_{\alpha_{i_{N+1}}}^\dag}{P(\alpha_{i_{N+1}}|\alpha_{i_1}\ldots\alpha_{i_N})\cdot\,J(\alpha_{i_1},\ldots,\alpha_{i_N})}\nonumber\\
              &=&\,\frac{M_{\alpha_{i_{N+1}}}\ldots\,M_{\alpha_{i_1}}\theta_0\,M_{\alpha_{i_1}}^\dag\ldots\,M_{\alpha_{i_{N+1}}}^\dag}{J(\alpha_{i_1},\ldots,\alpha_{i_{N+1}})}
\eea
Where we have used Bayes theorem in the last step. Therefore, eqn.(\ref{eq:stateNpp}), assumed valid for N, is true for $N+1$ also. On
using $Tr\,\theta_{N+1}\,=\,1$, it follows that
\be
\label{jointN+1}
J(\alpha_{i_1},\ldots,\alpha_{i_{N+1}})\,=\,Tr\,M_{\alpha_{i_{N+1}}}\ldots\,M_{\alpha_{i_1}}\theta_0\,M_{\alpha_{i_1}}^\dag\ldots\,M_{\alpha_{i_{N+1}}}^\dag
\ee
thus showing that the eqn.(\ref{eq:jointN}), assumed valid for N, is also valid for $N+1$. Since they are valid, by construction, for $N=1,2$,
they are valid for all values of n.

\section{QND measurements}
 So far we have not specified any particular form(s) for the measurement interaction $U$; any arbitrary measurement interaction would have 
been 
fine. For example, the measurement interactions of typical von Neumann models depending on {\it only one} system observable, or, the 
Arthurs-Kelley measurement interaction involving two non-commuting system observables \cite{arthurskelly1965}.

That brings us to the notion of Quantum Non-demolition(QND) measurements already alluded to in the introduction. Very generally speaking, these
measurements leave at least one of the system states {\it undisturbed}. In other words, the mapping introduced earlier must have {\it Fixed
Points}. Determining the fixed points for given initial states and measurement interaction in all generality is a difficult technical
problem. See \cite{quopfixed2002} for some results. For a very detailed, conceptual and technical exposition of QND measurements, see
\cite{brakha1992}.

However, when the measurement interaction is of the form $U({\hat q},{\hat Q})$ involving a single system observable ${\hat q}$, it is
straightforward to show that the measurements are of QND type, and furthermore, that the fixed points are just the eigenstates $|q_i\rangle$
of ${\hat q}$. This is also equivalent to another frequently used form of the QND criterion
\be
\label{eq:qndcomm}
[U,{\hat q}]\,=\,0
\ee
One may be tempted to interpret this to mean $U,{\hat q}$ can be simultaneously diagonalized but as these two operate on different Hilbert
spaces such an interpretation is not very precise. We will give a more precise meaning in terms of the measurement operators.
Introducing the eigenvalues and eigenstates of the system observable ${\hat{q}}$
\be
\label{eq:observablespectrum}
{\hat{q}}|q_i\rangle\,=\,q_i|q_i\rangle
\ee
Consider the action of the measurement operator $M_{\alpha_I}$ on $|q_i\rangle$:
\be
\label{eq:measopeigenf}
M_{\alpha_I}|q_i\rangle=\langle\,\alpha_I|U({\hat{q}},{\hat{Q}})|\phi_0\rangle|q_i\rangle\,=\,\lambda_I^i\,|q_i\rangle
\ee
where
\be
\label{eq:measopeigenv}
\lambda_I^i\,=\,\langle\,\alpha_I|U(q_i,{\hat{Q}})|\phi_0\rangle
\ee
These can be combined into a spectral decomposition for $M_{\alpha_I}$:
\be
\label{eq:measopspect}
M_{\alpha_I}\,=\,\sum_i\,\lambda_I^i\,|q_i\rangle\langle\,q_i|
\ee 
This shows that $M_{\alpha_I}$ and ${\hat q}$ are simultaneously diagonalizable with $\lambda_I^i$ being the eigenvalue of $M_{\alpha_I}$
corresponding to the simultaneous eigenstate $|q_i\rangle$. This can be rephrased as a more useful criterion for QND, in place of
eqn.(\ref{eq:qndcomm}):
\be
\label{eq:measqnd}
[M_{\alpha_I},{\hat q}]\,=\,0
\ee
On using eqn.(\ref{eq:measmap}) and eqn.(\ref{eq:measopeigenv}) it is easy to see that the density matrices $|q_i\rangle\langle\,q_i|$ are
fixed points of the map for every $i$ thus explicitly demonstrating the QND nature of the measurements.
\subsection{Important Properties of $\lambda_I^i$}
\label{subsec:lambdaprops}
Now we establish a number of important properties of $\lambda_I^i$ that are crucial for our work. To begin with 
\be
\label{eq:lambprops1}
\sum_I\,M_{\alpha_I}^\dag\,M_{\alpha_I}={\bf{I}} \rightarrow \sum_I\,|\lambda_I^i|^2\,=\,1 .
\ee
This holds for every $i$.
This points to $|\lambda_I^i|^2$ being some sort of probability. Indeed, as can be  verified easily by considering $M_{\alpha_I}|q_i\rangle$,
$|\lambda_I^i|^2$ is the probability of obtaining the outcome $\alpha_I$ when measured on the pure state $|q_i\rangle$:
\be
\label{eq:lambprob}
P(\alpha_I|i)\,=\,Tr\, M_{\alpha_I}\,|q_i\rangle\langle\,q_i|M_{\alpha_I}^\dag\,=\,|\lambda_I^i|^2
\ee
We use the expression for $\lambda_I^i$:
\be
\label{eq:lambexp}
\lambda_I^i\,=\,\langle\,\alpha_I|U(q_i,{\hat{Q}})|\phi_0\rangle
\ee 
to infer two further properties. When, the observable ${\hat q}$ has {\it degenerate spectrum} i.e. $q_i=q_j$ for some $i,j$.
\be
\label{eq:lambdeg}
|\lambda_I^i|^2\,=\,|\lambda_I^j|^2\quad \forall\, I
\ee
The situation is more complex when the eigenstates $|q_i\rangle,|q_j\rangle$ are {\it non-degenerate} i.e. $q_i\ne\,q_j$. In this case,
\be
\label{eq:lambndeg}
|\lambda_I^i|^2\ne\,|\lambda_I^j|^2
\ee
for {\it at least} one I.
This needs some explanation. For measurements to be considered adequate, they should be able to distinguish any pair of orthogonal states, at
least orthogonal pure states. Such states are as distinct as possible in quantum theory. Therefore, the statistics of measurements for
such orthogonal states must be distinct. In the present context, such measurement statistics are coded in $|\lambda_I^i|^2$ and therefore,
for the orthogonal eigenstates $|q_i\rangle$ and $|q_j\rangle$ for $i\ne\,j$, their mesasurement statistics must differ for at least one I.
Note that it is not necessary for the statistics to differ for every I.

In our analyses later on, it will be useful to introduce
\be
\label{eq:muij}
\mu_{ij}\,=\,\sum_I\,|\lambda_I^i||\lambda_I^j|
\ee
Let us consider 
\be
\label{eq:muineq}
\sum_I\,(|\lambda_I^i|\,-\,|\lambda_I^j|)^2\,\ge\,0
\ee
On using the definition of $\mu_{ij}$,
\be
\label{eq:muineq2}
\sum_I\,(|\lambda_I^i|\,-\,|\lambda_I^j|)^2\,=\,2(1\,-\,\mu_{ij})\,\ge\,0\rightarrow\,\mu_{ij}\,\le\,1
\ee
We can recast the properties of $\lambda_I^i$ derived earlier as: 
\be
\label{eq:mudiag}
\mu_{ii}=1 \quad \forall i
\ee
The cases when $i\ne\,j$ have to be analysed separately for when $i,j$ correspond to i) non-degenerate states, and, ii) when they
correspond to degenerate states. For the non-degenerate case, it follows that
\be
\label{eq:munondeg}
\mu_{ij}\,<\,1.
\ee
On the other hand, when they correspond to degenerate states,
\be
\label{eq:mudeg}
\mu_{ij}\,=\,1
\ee
The importance of these will become clear shortly.
\subsection{Repeated QND mesurements}
In this subsection, we shall explicitly evaluate all the important results of sec.(\ref{sub:genrepeat}) for repeated QND measurements. The
starting point is the spectral relation of eqn.(\ref{eq:measopspect}):
$$
M_{\alpha_I}\,=\,\sum_i\,\lambda_I^i\,|q_i\rangle\langle\,q_i|
$$
Here $|q_i\rangle$ are the complete set of eigenstates of the observable ${\hat q}$. Introducing the projectors $\Pi_i\,=\,|q_i\rangle\langle\,q_i|$, this can be written equivalently as
\be
\label{eq:measopspect2}
M_{\alpha_I}\,=\,\sum_i\,\lambda_{\alpha_I}^i\,\Pi_i
\ee
The projectors $\Pi_i$ satisfy
\be
\label{eq:projprop}
\Pi_i\,\Pi_j\,=\,\delta_{ij}\,\Pi_j\quad\quad\, \sum_i\,\Pi_i\,=\,{\bf I}\quad\quad Tr\,\Pi_i\,=\,1
\ee
An important consequence of eqn.(\ref{eq:measopspect2}) is the commutativity relation
\be
\label{eq:projmeasopcomm}
[M_{\alpha_I},\Pi_i]\,=\,0\quad\quad \forall I,i
\ee
\subsection{Degeneracy classes}
\label{subsec:degenclass}
We have so far not addressed the issue of whether some of the eigenstates belong to degenerate subspaces. We do that now. 
Let us label the different subspaces by ${\bar{i}}$, and let $d_{{\bar{i}}}$ denote their dimensionalities, which are just the degrees
of their degeneracies. We shall let the classes include even the non-degenerate cases, for which $d_{{\bar{i}}}$ is just 1.
Let us also label an arbitrarily chosen orthonormal basis for the
subspace by $({\bar{i}},i_0)$ with $i_0$ taking values $1,2,\ldots,d_{\bar{i}}$. The degeneracy of the eigenstates spanning the subspace is
reflected in $q_{{\bar{i},i_0}}\,=\,q_{\bar{i}}$, for all $i_0$. As the eigenvalues $\lambda_{\alpha_I}^i$ of the measurement
operators $M_{\alpha_I}$ only depend on the eigenvalues of ${\hat q}$, it also follows that $\lambda_{\alpha_I}^{{\bar{i}},i_0}\,=\,
\lambda_{\alpha_I}^{{\bar{i}}}$ for every $\alpha_I$. The summation in eqn.(\ref{eq:measopspect2}) can be replaced by a double summation
over ${\bar{i}}$ and $i_0$.
\bea
\label{eq:measspectdeg}
M_{\alpha_I}\,&=&\,\sum_{{\bar{i}},i_0}\,\lambda_{\alpha_I}^{{\bar{i}}}\,\Pi_{{\bar{i}},i_0}\nonumber\\
                          &=&\,\sum_{\bar{i}}\,\lambda_{\alpha_I}^{{\bar{i}}}\,\sum_{i_0}\,\Pi_{{\bar{i}},i_0}
\eea
On introducing the projector $\Pi_{{\bar{i}}}$ for the degenerate subspace, given explicitly by
\be
\label{eq:degproj}
\Pi_{{\bar{i}}}\,=\,\sum_{i_0=1}^{d_{{\bar{i}}}}\,\Pi_{{\bar{i}},i_0}
\ee
eqn.(\ref{eq:measspectdeg}) can be rewritten as
\be
\label{eq:measspectdeg2}
M_{\alpha_I}\,=\,\sum_{{\bar{i}}}\,\lambda_{\alpha_I}^{{\bar{i}}}\,\Pi_{{\bar{i}}}
\ee
which formally looks similar to eqn.(\ref{eq:measopspect2}) but with summation over just ${\bar{i}}$. The projectors $\Pi_{{\bar{i}}}$ for
the entire subspace labelled by ${\bar{i}}$ satisfy properties very imilar to eqn.(\ref{eq:projprop}) for one-dimensional projectors $\Pi_i$:
\be
\label{eq:degprojprop}
\Pi_{{\bar{i}}}\,\Pi_{{\bar{j}}}\,=\,\delta_{{\bar{i}}{\bar{j}}}\,\Pi_{{\bar{i}}}\quad\quad\, \sum_{\bar{i}}\,\Pi_{\bar{i}}\,=\,{\bf I}\quad\quad Tr\,\Pi_{{\bar{i}}}\,=\,d_{{\bar{i}}}
\ee
only the trace condition changes.

As in the case of fully non-degenerate spectrum, the eigenvalues of the projectors $\Pi_{\bar{i}}$ are still 0 and 1. On the degenearate
subspace ${\bar{i}}$ the eigenvalues are 1, and 0 otherwise. THere are exactly $d_{\bar{i}}$ eigenvalues equalling 1. This is also reflected
in the trace condition. A useful way of expressing this is that on the degenerate subspace $\Pi_{\bar{i}}$ is just the $d_{\bar{i}}\times\,d_{\bar{i}}$ unit matrix.
The analog of eqn.(\ref{eq:projmeasopcomm}) is now given by
\be
\label{eq:degprojmeasopcomm}
[M_{\alpha_I},\Pi_{{\bar{i}}}]\,=\,0\quad\quad \forall I,{\bar{i}}
\ee

Consider the following relation for the non-degenerate case:
\be
\label{eq:projeigenndeg}
M_{\alpha_I}\,\Pi_j\,=\,\sum_i\,\lambda_{\alpha_I}^i\,\Pi_j\Pi_i\,=\,\lambda_{\alpha_I}^j\,\Pi_j
\ee
where we used eqn.(\ref{eq:measopspect}), as well as eqn.(\ref{eq:projprop}). This can be viewed as a generalized eigenvalue equation
where the eigenstates are projection operators.

A generalization of this for the projectors $\Pi_{{\bar{i}}}$ whose dimensionalities are not restricted to be 1 is easy to prove:
\be
\label{eq:projeigendeg}
M_{\alpha_I}\,\Pi_{{\bar{j}}}\,=\,\sum_{{\bar{i}}}\,\lambda_{\alpha_I}^{{\bar{i}}}\,\Pi_{{\bar{j}}}\Pi_{{\bar{i}}}\,=\,\lambda_{\alpha_I}^{{\bar{j}}}\,\Pi_{{\bar{j}}}
\ee
In view of the earlier remarks, this means that the QND measurement operators act essentially as identity operators on the degenerate subspace.
More precisely they are $\lambda_I^{\bar{i}}\cdot{\bf I}_{\bar{i}}$.

Using these repeatedly it is easy to show
\be
\label{eq:measoprepeatprojndeg}
M_{\alpha_{i_2}}\,M_{\alpha_{i_1}}\,=\,\sum_i\,\lambda_{\alpha_{i_1}}^i\lambda_{\alpha_{i_2}}^i\,\Pi_i
\ee
and its generalization to include degenerate subspaces
\be
\label{eq:measoprepeatprojdeg}
M_{\alpha_{i_2}}\,M_{\alpha_{i_1}}\,=\,\sum_{{\bar{i}}}\,\lambda_{\alpha_{i_1}}^{{\bar{i}}}\lambda_{\alpha_{i_2}}^{{\bar{i}}}\,\Pi_{{\bar{i}}}
\ee
These can be easily extended to products of arbitrary number of measurement operators:
\be
\label{eq:measoprepeatnprojndeg}
M_{\alpha_{i_n}}\dots\,M_{\alpha_{i_1}}\,=\,\sum_i\,\prod_{j=1}^n\,\lambda_{\alpha_{i_j}}^i\,\Pi_i
\ee
and,
\be
\label{eq:measoprepeatnprojdeg}
M_{\alpha_{i_n}}\ldots\,M_{\alpha_{i_1}}\,=\,\sum_{{\bar{i}}}\,\prod_{j=1}^n\,\lambda_{\alpha_{i_j}}^{{\bar{i}}}\,\Pi_{{\bar{i}}}
\ee
Using these it is easy to see
\be
\label{eq:thetanproj}
M_{\alpha_{i_n}}\ldots\,M_{\alpha_{i_1}}\theta_0\,M_{\alpha_{i_1}}^\dag\ldots\,M_{\alpha_{i_n}}^\dag\,=\,\sum_{{\bar{i}},{\bar{k}}}\,\prod_{j=1}^n\,\lambda_{\alpha_{i_j}}^{{\bar{i}}}{\lambda_{\alpha_{i_j}}^{{\bar{k}}}}^*\,\Pi_{{\bar{i}}}\theta_0\Pi_{{\bar{k}}}
\ee
On noting
\be
\label{eq:thetantr}
Tr \Pi_{{\bar{i}}}\theta_0\Pi_{{\bar{k}}}\,=\,\delta_{{\bar{i}}{\bar{k}}}\,Tr\,\theta_0\,\Pi_{{\bar{i}}}\,=\,\delta_{{\bar{i}}{\bar{k}}}\,|c_{{\bar{i}}}|^2
\ee
with 
\be
\label{eq:degoverlap}
|c_{{\bar{i}}}|^2\,=\,Tr\,\Pi_{{\bar{i}}}\theta_0
\ee
denoting the probability of finding the subspace ${{\bar{i}}}$ in the initial state $\theta_0$, the joint probability
distribution of eqn.(\ref{eq:jointN}) is now given by
\be
\label{eq:jointnfin}
J(\alpha_{i_1}\ldots,\alpha_{i_n})\,=\,\sum_{{\bar{i}}}\,\prod_{j=1}^n\,|\lambda_{\alpha_{i_j}}^{{\bar{i}}}|^2\,|c_{{\bar{i}}}|^2
\ee
The conditional probabilities $P(\alpha_{i_n}|\alpha_{i_1},\ldots,\alpha_{i_{n-1}})$ can be easily computed on using eqn.(\ref{eq:jointnfin})
and the Bayes theorem. We shall not write them down explicitly as their explicit forms will not be made use of later.

The explicit expression for $\theta_n$ is given by
\be
\label{eq:thetanfin}
\theta_n\,=\,\frac{\sum_{{\bar{i}}{\bar{k}}}\,\prod_{j=1}^n\,\lambda_{\alpha_{i_j}}^{{\bar{i}}}\lambda_{\alpha_{i_j}}^{{\bar{k}}}\,\Pi_{{\bar{i}}}\theta_0\Pi_{{\bar{k}}}}{J(\alpha_{i_1},\ldots,\alpha_{i_n})}
\ee
As mentioned before, if the initial state $\theta_0$ is pure, all subsequent $\theta_n$ are also pure. Suppose the initial pure state $\theta_0\,|\psi\rangle\langle\,\psi|$ with
\be
\label{eq:inipure}
|\psi\rangle\,=\,\sum_i\,c_i|q_i\rangle
\ee
$\theta_n\,=\,|\psi_n\rangle\langle\,\psi_n|$, with $|\psi_n\rangle$(modulo an irrelevant phase) given by 
\be
\label{eq:psifinndeg}
|\psi_n\rangle\,=\,\frac{\sum_i\,\prod_{j=1}^n\,\lambda_{\alpha_{i_j}}^{i}\,c_i|q_i\rangle}{\sqrt{J(\alpha_{i_1},\ldots,\alpha_{i_n})}}
\ee
for the case when ${\hat q}$ has only a non-degenerate spectrum:
\subsection{Properties of $\lambda_{\alpha_I}^{{\bar{i}}}$}
\label{subsec:deglambprops}
In this brief subsection, we give the generalizations, without details, of the properties of $\lambda_{\alpha_I}^{{\bar{i}}}$ when the states 
are enumerated by their degeneracy classes. These follow from a straightforward application of eqn.(\ref{eq:measspectdeg2}), in place of the
corresponding eqn.(\ref{eq:measopspect2}) in terms of basis vectors. 

\be
\label{eq:lambpropdeg1}
\sum_{\alpha_I}\,M_{\alpha_I}^\dag\,M_{\alpha_I}={\bf{I}} \rightarrow \sum_{\alpha_I}\,|\lambda_{\alpha_I}^{{\bar{i}}}|^2\,=\,1 .
\ee
This holds for every ${\bar{i}}$.

As in eqn.(\ref{eq:muij}), we now introduce
\be
\label{eq:mubarij}
\mu_{{{\bar{i}}}{\bar{j}}}\,=\,\sum_{\alpha_I}\,|\lambda_{\alpha_I}^{{\bar{i}}}||\lambda_{\alpha_I}^{{\bar{j}}}|
\ee
Using arguments similar to the ones in eqn.(\ref{eq:muineq2}) it can be shown that for ${{\bar{i}}}\,\ne\,{\bar{j}}$,
\be
\label{eq:mubarineq}
\mu_{{\bar{i}}{\bar{j}}}\,\le\,1
\ee
We can recast the properties derived in eqn.(\ref{eq:lambpropdeg1}) as: 
\be
\label{eq:mubardiag}
\mu_{{\bar{i}}{\bar{i}}}=1 \quad \forall {\bar{i}}
\ee
The importance of these will become clear shortly.

\section{Generation of quantum trajectories}
Now we come to the trajectories, in ${\cal H}_S$, generated by repeated generalized measurements. The starting point is the system in some
unknown state $\theta_0$ and the probe in a suitably chosen pure state $|\phi_0\rangle\langle\,\phi_0|$. A generalised measurement is done on 
this resulting in a probe-outcome $\alpha_{i_1}$, the probe state in $|\alpha_{i_1}\rangle$ and the system state in $\theta_1$, as given by the
measurement map. After this the generalized measurement is repeated. Before doing so the probe state is restored to what one started with i.e.
$|\phi_0\rangle\langle\,\phi_0|$. This is to render identical the repeated measurements. But \cite{bauerlong2013} have shown that such
restrictions are really not necessary. The probe outcome after this is $\alpha_{i_2}$ with the system state being mapped from $\theta_1$ to
$\theta_2$ and the probe state is $|\alpha_{i_2}\rangle$. The probability distributions for the outcomes $\alpha_{i_1},\alpha_{i_2}\ldots$ are 
given by the {\it Joint Probability Distributions}
$J(\alpha_{i_1},\alpha_{i_2}\ldots)$. It is important to keep in mind, for example, the probability distribution for $\alpha_{i_2}$, which 
explicitly depends on $\theta_1$ which in turn depends on $\alpha_{i_1}$ is conditional on $\alpha_{i_1}$. 

Both the sequence of outcomes $\alpha_{i_1},\alpha_{i_2}...\alpha_{i_n}\ldots$ and the sequence of states $\theta_1,\theta_2\ldots$ define a 
{\it quantum trajectory} whose evolution is purely probabilistic. At the nth stage, when the system state sequence 
$\theta_1,\theta_2\ldots\theta_n$ is
fixed for a given trajectory, measurement can give rise to all possible values for the next outcome $\alpha_{i_{n+1}}$ with probability
distribution given by the conditional probability distribution $P(\alpha_{i_{n+1}}|\alpha_{i_1},\alpha_{i_2}\ldots\alpha_{i_n})$ 
The usefulness of both these sequences for gleaning the unknown initial state will be the main
focus of the remainder of this paper.
\section{Gaussian QND Measurements}
\label{sec:gaussianpovm}
In this section we shall illustrate many of ideas discussed so far with what we call {\it Gaussian QND} measurements. These are characterized
by measurement operators that are gaussian in the outcomes and in addition satisfy the QND criterion. We shall separately  analyse i) a single
measurement on an ensemble of identically prepared states, generic in quantum mechanics, as well as, ii) repeated measurements on a
single copy. 

This is a summary of \cite{weakrepeat2014}. In view of the general formalism already presented in the earlier sections, we shall not present
the explicit derivations given in the 2014 paper. The  pointer variable is taken to be the continuous variable ${\hat P}$, the momentum of 
the probe. The corresponding pointer states are taken to be the momentum eigenstates $|p\rangle$.

As is well known, such states obey continuum normlization conditions, which apart from being mathematically not so well-defined, lead to
various difficulties in physical applications also. In our current discussion of generalized measurements so far, the pointer states had been 
taken to be discrete, without such difficulties. In particular, the initial probe state $|\phi_0\rangle$( taken to be pure for simplicity) 
could have been any of the discrete pointer states, or their superposition. But with continuous pointer states, things are more nuanced! 
In practice, the initial probe states are taken to be dwnarrow gaussian wave packets in momentum representation:
\be
\label{eq:probewavepacket}
|\phi\rangle_0\,=\,N\,\int\,dp\,e^{-\frac{p^2}{2\Delta^2}}|p\rangle\quad\quad\,N^2\sqrt{\pi\Delta^2}=1
\ee
The magnitude of $\Delta$ determines whether the measurements are weak or close to being projective. The former are realised when $\Delta$
is very large, and the latter for very small values.

For simplicity, the initial state of the system is also taken to be pure i.e. $\theta_0\,=\,|\psi\rangle\langle\,\psi|$, with
\be
\label{eq:sysiniquad}
|\psi\rangle = \sum_i\,c_i\,|q_i\rangle\quad\quad \sum_i\,|c_i|^2=1
\ee
where $|q_i\rangle$ are the eigenstates of the system observable ${\hat q}$ i.e.
\be
\label{eq:sysobsquad}
{\hat q}|q_i\rangle\,=\,q_i|q_i\rangle
\ee
The measurement interaction was taken to be of an impulsive kind involving ${\hat q}$S and the position operator ${\hat Q}$(canonically 
conjugate to ${\hat P}$ as per the von Neumann model. We refer the reader to \cite{weakrepeat2014} for all the details. It suffices to
give the resulting measurement operators $M_p$:
\be
\label{eq:measopquad}
M_p =  N\,\sum\,e^{-\frac{(p-q_i)^2}{2\Delta^2}}|q_i\rangle\langle q_i|
\ee
The corresponding eigenvalues of $M_p$, the $\lambda_p^i$ are given by
\be
\label{eq:lambdaquad}
\lambda_p^i\,=\,N\,e^{-\frac{(p-q_i)^2}{2\Delta^2}}
\ee
As already discussed, $|\lambda_p^i|^2$ is the probability distribution for obtaining the outcome $p$ when measurements are done on the
system in $|q_i\rangle$. The mean of this distribution is at $q_i$ with a variance $\frac{\Delta}{\sqrt{2}}$. 

It is easy to see that they satisfy the constraints on measurement operators:
\be
\label{eq:measconstraintquad}
\int\,dp\,M_p^\dag\,M_p\,=\,\sum_i\,|q_i\rangle\langle\,q_i|\,=\,{\bf I}
\ee
\subsection{Single measurements on ensemble of states}
Applying our general results it is easy to see that The probability $P(p)$ of obtaining the probe outcome $p$ is given by:
\be
\label{eq:quadprob1}
P(p)\,=\,\langle\,\psi|M_p^\dag\,M_p|\psi\rangle\,=\,N^2\,\sum_i\,|\alpha_i|^2\,e^{-\frac{(p-q_i)^2}{\Delta^2}}
\ee
It is not possible to explicitly carry out the summation above. However, the first and second moments can be calculated, and they are
sufficient for the analysis to be presented shortly. The first moment is given by
\be
\label{eq:quadmean}
\langle\,p\,\rangle_\psi\,=\,\int\,dp\,p\,P(p)\,=\,\sum_i\,|c_i|^2\,q_i\,=\,\langle\,\psi|{\hat q}|\psi\rangle
\ee
This is just the expectation value of the system observable in the unknown initial state $|\psi\rangle$. By repeating the measurements with an
optimal set of system observables, the unknown state can be determined. It is in this sense that the statistics of even weak measurements
has full information about the unknown initial state. Actually, since this conclusion is independent of the value of $\Delta$, it is
true for arbitrary generalized measurements of this type.

The second moment is also easily calculated:
\be
\label{eq:quadvar}
\langle\,p^2\,\rangle_\psi\,=\,\frac{\Delta^2}{2}\,+\,\sum_i\,|c_i|^2\,q_i^2
\ee
For weak measurements $\Delta^2$ will be seen to be very large and the second moment is $\frac{\Delta^2}{2}$ to a very good approximation.
The state vector $|\psi\rangle_f$ after the first measurement is given by
\be
\label{eq:gaussianpost2}
|\psi_f\rangle\,=\,\frac{\sum_i\,\alpha_i\,e^{-\frac{(p-q_i)^2}{2\Delta^2}}|q_i\rangle}{\sqrt{\sum_i\,|\alpha_i|^2\,e^{-\frac{(p-q_i)^2}{\Delta^2}}}}
\ee
When $\Delta$ is very small, approaching zero, the measurement operators become essentially delta functions:
\be
\label{eq:measopproj}
M_p\xrightarrow[\Delta\rightarrow\,0]\,\sum_i\,\delta(p-q_i)|q_i\rangle\langle\,q_i|
\ee
bringing the measurements closer and closer to projective measurements. The outcomes become increasingly correlated with the eigenvalues of
the measured observable, and the post-measurement state with the corresponding eigenstate.

 At the other extreme is the limit $\Delta\rightarrow\,\infty$. It is easy to see that in this limit, for $|p|<<\Delta$, the measurement
operators approach constant multiple of identity:
\be
\label{eq:measopweak}
M_p\xrightarrow[\Delta\rightarrow\,\infty]\,N\,\sum_i|q_i\rangle\langle\,q_i|=N\,{\bf I}
\ee
 This means for such values of outcomes, the initial state hardly changes( N cancels out between the numerators and the denominators in 
eqn.(\ref{eq:gaussianpost2})), and for this reason this regime is called {\it weak measurements. In other words, in this regime the measurement
map becomes very close to the identity map.} The factor $N$ vanishes in this weak limit
signifying a very low probability for such cases. It is of course true that for $|p|\ge\,\Delta$ there are significant changes. But such values are exponentially improbable as per eqn.(\ref{eq:quadprob1}).

 Though we have shown these for pure  initial states, it is obvious that they hold for arbitrary initial states. For generalized(POVM)
measurements, the second moments of $|\lambda_I^i|^2$ play the same role as $\Delta^2$(see sec.(\ref{sec:genensemble}).

It is to be expected that when state changes are small during measurements, information obtained will also be small. So can one get any 
information about the original state from the statistics of outcomes of repeated weak measurements? Because of the gaussian forms of $P(p)$ all
moments of that probability distribution can be analytically evaluated, irrespective of whether $\Delta$ is small or big. Let us compute the
first moment of P(p):
\be
\label{eq:quadmean2}
\langle\,{\hat{p}}\,\rangle_\psi\,=\,\mu\,=\,\int\,dp\,p\,P(p)\,=\,\sum_i\,|\alpha_i|^2\,q_i\,=\,\langle\,\psi|{\hat q}|\psi\rangle
\ee
This shows that the statistics of outcomes can fully determine the expectation value of the observable ${\hat q}$ in the unknown original
state, much like the projective measurements! Repeating the mesurements for an optimal set of observables, the state $|\psi\rangle$ itself
can be fully determined. While this sounds encouraging, the second moment, rather the variance, tells the flip side of weak measurements!
\be
\label{eq:quadvar2}
(\Delta\,p)_\psi^2=\,\frac{\Delta^2}{2}\,+\,(\Delta\,q)_\psi^2
\ee
With $\Delta$ being very large for weak measurements, these measurements are also much noisier, demanding very large ensembles to keep
statistical errors under control.

The average of M outcomes $p_1,p_2,\ldots,p_M$ given by $y_M\,=\,\frac{\sum_{i=1}^M\,p_i}{M}$ (the bin-average) is a particularly useful
quantity to study how the true average $\langle\,p\,\rangle$ is approached by actual realisations. In particular, the probability distribution
for $y_M$, in terms of the joint probability distribution $J(p_1,\ldots,p_M)$ is  given by
\be
\label{eq:binavprob}
P(y_M)\,=\,\int\,\prod_{j=1}^M\,dp_j\,\delta(y_M-\frac{\sum_i\,p_i}{M})\,J(p_1,\ldots,p_M)
\ee
For single measurements on ensemble of identically prepared states, $p_i$ are independently and identically distributed. Hence
\be
\label{eq:quadensjoint}
J(p_1,\ldots,p_M)\,=\,\prod_{j=1}^M\,P(p_j)
\ee
Though the probability distribution $P(p)$ can not be evaluated explicitly, the central limit theorem, which only requires knowledge of the
first two moments of $P(p)$, allows one to explicitly evaluate $P(y_M)$ to be
\be
\label{eq:quadensprobym}
P(y_M)\,=\,\sqrt{\frac{M}{\pi\Delta^2}}\,e^{-\frac{M\,(y_M-\mu)^2}{\Delta^2}}
\ee
This shows that as M becomes very large $y_M$ approaches the true average $\mu$ with a variance $\frac{\Delta}{\sqrt{2M}}$ decreasing as 
$\sqrt{M}$, a standard result in data analysis.

\subsection{Repeated POVM measurements on a single copy}
The fact that statistics of outcomes has information about the unknown original state gives rise to the hope that the statistics of a very
large number of repeated POVM measurements, but on a single copy of the unknown state, may likewise determine the unknown initial state.
Should one succeed in that, ensembles of identically prepared states would not be a necessity in quantum theory! At least not very large 
ensembles. Just as many copies would suffice as the number of observables whose expectation values determine the states of the system. We now 
turn to an analysis of this foundational issue within the gaussian QND measurement schemes.

In the previous subsection we analysed single such measurements performed on an ensemble of states and we considered the probability 
distribution $P(y_M)$ where $y_M$ was the average over M measurement outcomes. The outcomes over which this averaging was done could
have been any collection of outcomes. But now we want to take this set to consist of, say, the first M outcomes along a given trajectory. The
trajectories are all generated by the action of M repeated measurements, all on a single copy of the system in an unknown state. 

The question we wish to address whether this sequential data along one given trajectory can be used to determine the original unknown
state. But the conceptual difficulty is that any given trajectory is one particular realisation of the random processes generating the 
trajectories. To one of us(NDH), in 2014, it didn't seem possible, even in principle, to analyse a single trajectory.. Instead, he wanted 
to carry out averaging over all possible
outcomes of all the repeated measurements and view that as a double averaging, namely, first averaging along a given trajectory, and,then 
average that over all trajectories. Hoping that the second averaging did not wash out all the interesting features of the first averaging,
he hoped to glean enough information about the individual trajectory itself.

We start with explicit formulae.  From the general formalism given before, the joint probability distribution and the state of the 
system after n measurements are given by
\begin{eqnarray}
\label{eq:quadrepeat}
J(p_1,p_2,\ldots,p_n) &=& (N^{2n})\,\sum_i\,|c_i|^2\,\prod_{j=1}^n\,e^{-\frac{(p_j-q_i)^2}{\Delta^2}}\nonumber\\
|\psi_n\rangle &=& \frac{\sum_i\,\prod_{j=1}^n\,e^{-\frac{(p_j-q_i)^2}{\Delta^2}}c_i\,|q_i\rangle}{\sqrt{\sum_i\,\prod_j\,|c_i|^2\,e^{-\frac{(p_j-q_i)^2}{\Delta^2}}}}\nonumber
\end{eqnarray}
\subsection{Consequences}
The dramatic change now is reflected in the form of the joint probability distribution which is no longer expressible as a product of identical
probability distributions. In other words, the $p_i$ are no longer independedly distributed. One might think that Central Limit Theorem would
no longer be applicable. However, the joint probability distribution is now a sum of terms each of which has a factorizable form. For
generalized measurements, to be discussed soon, the central limit theorem can be applied to each term. But for the gaussian measurement
operators under consideration, one need not even seek the power of CLT! All integrations can be carried out explicitly.

As before, let us study $y_M$, the average of the M outcomes. It is to be noted that eqn.(\ref{eq:binavprob}) continues to be valid even now
as long as we use the joint probability distribution given by eqn.(\ref{eq:quadrepeat}). It is straightforward to prove that
\begin{equation}
\label{eq:repeatprobym}
P(y_M) = \sqrt{\frac{M}{\pi\Delta_p^2}}\,\sum_i\,|c_i|^2\,e^{-\frac{(y_M-s_i)^2M}{\Delta_p^2}}
\end{equation}
The average value ${\bar y}_M$ is also easily calculated:
\be 
\label{eq:quadrepeatymav}
{\bar y}_M\,=\,\int dy_M\,y_M\,P(y_M)\,=\, \sum_i\,|c_i|^2\,q_i\,=\,\mu
\ee
Likewise, the variance in $y_M$ can be shown to be $\frac{\Delta}{\sqrt{2M}}$. 

Both these are exactly the same that were found for single measurements on an ensemble of states. But the distributions $P(y_M)$ of 
eqn.(\ref{eq:quadensprobym}) and eqn.(\ref{eq:repeatprobym}) are dramatically different. To bring out these differences more vividly, let
us look at their asymptotic limits as $M\rightarrow\infty$:
\be
\label{eq:quadensasymp}
P_{ens}(y_M)\,\rightarrow\, \delta(y_M-\mu)\quad\quad \mu\,=\,\sum_i\,|c_i|^2\,q_i
\ee
while
\be
\label{eq:quadrepeatymasymp}
P_{repeat}(y_M)\,\rightarrow \sum_i\,|c_i|^2\,\delta(y_M-s_i)
\ee
Thus, unlike in the case of ensemble measurements(both strong and
weak), the distribution of $y_M$ is no longer peaked at the true average, with errors decreasing as $M^{-1/2}$. Instead, it is a weighted sum of sharp
distributions peaked around \emph{the eigenvalues}, exactly as in the strong measurement case. 

It is very important to notice that the asymptotic behaviour only depended on $M\rightarrow\,\infty$ {\it irrespective} of whether $\Delta$
itself was large or small. In other words, though the original question was motivated by the weakness of measurements, the conclusion itself
is general. We shall find the same later on when we address the asymptotic behaviour of quantum trajectories by more precise methods. The
meaning of this rather surprising result will be commented upon later.

At this point, nothing requires all the $p_i$ to lie along any trajectory. In fact, the averaging is done over all possible values of them.
However, the above results can be consistently interpreted in the following way:averaging over all possible $p_i$ can be thought of as a 
two step averaging. The first averaging is over $p_i$ along a particular trajectory, and the subsequent one as averaging the first average
over all possible trajectories.

The forms of eqns.(\ref{eq:repeatprobym},\ref{eq:quadrepeatymasymp}) then suggest that along each trajectory, ${\bar y}_M$ approaches
some eigenvalue, randomly, and hence the trajectories can be labelled by the eigenvalues $q_i$, or, by $i$ for short. The second averaging, 
namely, averaging over distinct trajectories, is then simply given by the probability of realising a particular trajectory. Again from the 
form of the above equations it follows that $|c_i|^2$ is that probability.

Remarkably, the situation has now become like that in projective measurements! The outcomes are in one-one correspondence with the
eigenvalues of the observable, and, the probability of any particular outcome is given by the Born rule!
It then follows that averages along a given trajectory can not give any information about the initial state of the single copy! Ensembles
again become inevitable. The other consequence is that a very large number of repeated POVM measurements on a single copy has the same invasive effect as a strong measurement.

So far we left open the possibility of the observable having degenerate spectrum, though we did not address it explicitly. It is quite
straightforward to do that. As before, we replace the summation over the eigenstates labelled by $i$ by summation over $({\bar{i}},i_0)$
with ${\bar{i}}$ being the label for the degeneracy class and $i_0$ the labelling of the orthonormal basis spanning the subspace. The
eigenvalues satisfy $s_{{\bar{i}},i_0}\,=\,s_{{\bar{i}}}$. It is then easy to recast the earlier equations as
\begin{equation}
\label{eq:repeatprobymdeg}
P(y_M) = \sqrt{\frac{M}{\pi\Delta_p^2}}\,\sum_{{\bar{i}}}\,|c_{{\bar{i}}}|^2\,e^{-\frac{(y_M-s_{{\bar{i}}})^2M}{\Delta_p^2}}
\end{equation}
and
\be
\label{eq:quadrepeatymasympdeg}
P_{repeat}(y_M)\,\rightarrow \sum_{{\bar{i}}}\,|c_i|^2\,\delta(y_M-s_{{\bar{i}}})
\ee
with
\be
\label{eq:cibar}
|c_{{\bar{i}}}|^2\,=\,\sum_{i_0}\,|c_{{\bar{i}},i_0}|^2
\ee
having the meaning of the probability of finding the degenerate subspace ${{\bar{i}}}$ in the initial state. Now the trajectories are
labelled by the degeneracy classes and their probabilities given by the corresponding generalization of the Born rule. As is well known, even
when all the non-degenerate eigenstates are fixed, there is infinite freedom in the choice of the degenerate eigenstates labelled by $i_0$
related to each other by $SU(d_{{\bar{i}}})$ transformations. The results so far obtained are all invariant under such transformations, as they
should be.

Two rather serious shortcomings of the analysis above are: i) the asymptotic state of the system along a given trajectory is not determined.
In other words, the von Neumann state reduction does not follow automatically, and, ii) while the above interpretation is plausible, it
does not follow rigorously. A more sophisticated analysis is needed to investigate individual trajectories. This will form the main body of 
the rest of this paper. It is however gratifying that that rigorous analysis upholds the essence of the interpretation proposed here.

On the basis of these considerations, the following picture emerges.The trajectories fall into families with members being labelled by the
eigenvalues of the observable ${\hat q}$, for both the degenerate and non-degenerate parts of the spectrum.
\begin{figure}[htp!]
  \centering
  \includegraphics[width=1.5in]{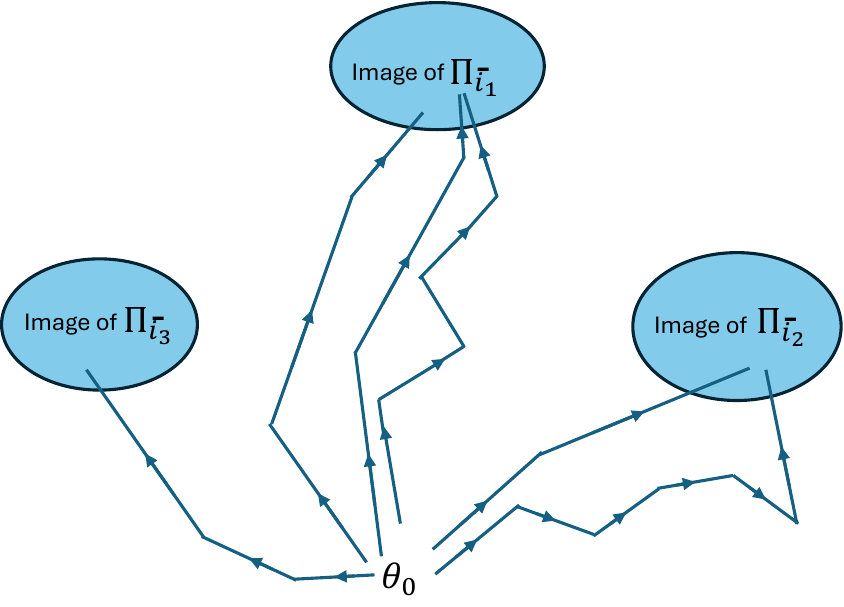}
  \label{}
\end{figure}
The probabilities for the trajectories are governed by Born's rule. As the measurements under consideration are of the QND type, the 
eigenstates $|i\rangle$ are the {fixed points} of the map generating the trajectories. However, as is characterstic of such maps, fixed points
are never really reached but only asymptotically approached.  As the number of POVM measurements becomes very large, the trajectory in state 
space hovers around a particular eigenstate.
\section{Generalized ensemble analysis}
\label{sec:genensemble}
One may wonder if the derivations and interpretations given above are artefacts of the gaussianity of the measurement operators, and
whether they would still persist for the generalized measurements. One feature will certainly disappear and that is the ability to do
the $p_i$-integrations explicitly without recourse to CLT.

For generalized QND measurements, as can be seen from eqn.(\ref{eq:jointnfin}), what replace the gaussians are 
$\lambda_{\alpha_{i_j}}^{{\bar{i}}}$. The joint probability distribution is a weighted sum of the product of 
$|\lambda_{\alpha_{i_j}}^{{\bar{i}}}|^2$. To proceed further, it suffices if these generalized probability distributions have well defined 
first and second moments. Then the CLT can be applied to the products and one can obtain the analogs of the results for the gaussian
QND measurements. Some important differences still persist. Firstly, the first moments, denoted by, say, $\xi_{{\bar{i}}}$, will in
general not be the eigenvalues $q_i$ though they are determined by them. So the trajectories will now be labelled by $\xi_{{\bar{i}}}$

So unless the generalized measurements are so designed that $\xi_{{\bar{i}}}$ uniquely determine $q_{{\bar{i}}}$, the trajectories will
no longer be labelled by the eigenvalues alone. We also saw that for single QND gaussian measurements on ensemble of states, the average
of the probe outcomes equalled $\langle\psi|{\hat q}|\psi\rangle$ so the measurements done with sufficiently many observables could determine
the unknown initial state (see eqn.(\ref{eq:quadmean},\ref{eq:quadmean2})). This will no longer be true if the first moments of 
$|\lambda_{\alpha_{i_j}}^{{\bar{i}}}|^2$ are not $q_{{\bar{i}}}$. Another difference, though not as serious as the one with first moments is
that the variances could be ${\bar{i}}$-dependent. Alter and Yamomoto in their book \cite{alterbook2001} also make several observations
about these issues.
\section{Analysis of individual trajectories?}
\label{sec:qtrajmain}
While the above analysis and the picture emerging from it is certainly plausibile,  it is far from an acceptable proof. It still does not 
show a way of analysing {\it individual trajectories}. 
This is because individual trajectories are {\it specific} realization of a random process.  A remarkable paper 
by Maassen and K\"ummerer \cite{maassen2006} showed the light at the end of the tunnel!\footnote{We thank Dr. A.R. Usha Devi for bringing
this paper to our attention.}
They showed that under repeated generalized measurements, trajectories almost surely {\it purify}
in the sense that almost always trajectories asymptotically become {pure} states irrespective of the states they started in.
They also pointed out exceptional cases where this does not happen. We will also give specific examples when such purification can not
happen. Subsequently we found many works that have addressed the behaviour of individual trajectories. Belavkin \cite{belavkinmultivar1992,
belavkincmp1992} had also proved such purification but made use of stochastic calculus for that purpose. All three approaches discussed in this
paper specifically avoid using such stochastic methods, using instead discrete time approach. Maassen and K\"ummerer also used the
discrete time approach.

The key ingredient in Maassen and K\"ummerer's analysis is the use of the so called Nielsen identities \cite{nielsen2001}. They take the
form
\be
\label{eq:nielsenids}
\sum_i\,P(i)\,Tr\,{\theta_i^\prime}^m\,\ge\,Tr\, \theta^m
\ee
where $\theta$ is the initial state, $\theta_i^\prime$ the state to which it is mapped after a measurement outcome $i$, and $P(i)$ the
probability of the outcome. Nielsen proved these identities using {\it Majorization} techniques \cite{nielsen2001}. A proof of these
identities based on SDE were given in \cite{ushaparth2017}. Maassen and K\"ummerer 
recast these for quantum trajectories as
\be
\label{eq:trajnielsen}
\mathbb{E}[Tr\,\theta_{n+1}^m|\theta_1,\ldots,\theta_1]\,\ge\,Tr\,\theta_n^m\quad\quad\,\forall n,m
\ee
and hence concluded that $Tr\,\theta_n^m$ are {\it submartingales}. Using the {\it martingale convergence theorem}, they arrive at their
results on purification. For our analysis the Nielsen identities are not required at all though we too invoke martingales, submartingales 
and the martingale convergence theorem. We explain these concepts in the next section.
Martingales have also been used by others in addressing the state reduction issue but they are not ab initio derivations from the 
principles of quantum mechanics. Examples are \cite{adler2001,stockton2004,stockton2005}, though these were based on stochastic differential
equations.  Ours as well as the works of 
\cite{bauershort2011,bauerlong2013,amini2011}, based on the discrete time approach,  are direct consequences of quantum theory.
\section{All about Martingales}
\label{sec:martingales}
Consider a sequence of random variables $x_0,x_1,\ldots,x_n$. At each stage there is a probability distribution for them to take
new values. The sequence is a {\it martingale} if 
\be
\label{eq:martingale}
\mathbb{E}[x_{n+1}|x_n,x_{n-1},\ldots\,x_0]\,=\,x_n
\ee
Where $\mathbb{E}[a|b_1,\ldots,b_n]$ is the  {\it Conditional Expectation} of $a$ given $b_1,\ldots,b_n$.

The sequence is a {\it supermartingale} if 
\be
\label{eq:supermartingale}
\mathbb{E}[x_{n+1}|x_n,x_{n-1},\ldots\,x_0]\,<\,x_n
\ee
The sequence is a {\it subMartingale} if 
\be
\label{eq:submartingale}
\mathbb{E}[x_{n+1}|x_n,x_{n-1},\ldots\,x_0]\,>\,x_n
\ee

When the random variables are bounded, a remarkable and counter-intuitive result follows called the {\it martingale convergence theorem},
which states that (sub/super)martingales almost surely converge.

\section{The asymptotic behaviour of trajectories}
\label{sec:trajasymp}
We now address the central issue of this paper, namely, the asymptotic behaviour of quantum trajectories. Our proof makes use of the
martingale properties of various quantities along a given trajectory as well as the properties of the eigenvalues $\lambda_{\alpha_I}^i$
of the QND measurement operators detailed in sec.(\ref{subsec:lambdaprops}).
\subsection{Martingales of quantum trajectories}
As martingales will play a central role in establishing the desired asymptotic behaviour, in this subsection we shall establish the
martingale property of a number of quantities along a given trajectory. Let $\theta_n$ be the density matrix of the system at the nth map.
Let the probe outcome of the measurement at the next step be $\alpha_{i_{n+1}}$ and the new state be $\theta_{n+1;\alpha_{i_{n+1}}}$.

We have highlighted the explicit dependence of the state at the $n+1$-th step on the outcome of the probe. This had been left implicit
all along to avoid clutter in the mathematical expressions. 
\be
\label{eq:thetan+1}
\theta_{n+1;\alpha_{i_{n+1}}}\,=\,\frac{M_{\alpha_{i_{n+1}}}\,\theta_n\,M_{\alpha_{i_{n+1}}}^\dag}{P(\alpha_{i_{n+1}}|\alpha_{i_1},\ldots,\alpha_{i_n})}
\ee
We have already given explicit expressions for the conditional probabilities above. However, it is not necessary to use those in what follows.

We shall identify many quantities that will have the martingale property. We first investigate if the entire density matrix
can be a martingale.
\bea
\label{eq:thetamartingale}
\mathbb{E}[\theta_{n+1}|\theta_1,\ldots,\theta_n]\,&=&\,\sum_{\alpha_{i_{n+1}}}\,\theta_{n+1;\alpha_{i_{n+1}}}\,P(\alpha_{i_{n+1}}|\alpha_{i_1},\ldots,\alpha_{i_n})\nonumber\\
&=&\,\sum_{\alpha_{i_{n+1}}}\,M_{\alpha_{i_{n+1}}}\,\theta_n\,M_{\alpha_{i_{n+1}}}^\dag\,\ne\,\theta_n
\eea
Thus for generic system states, the density matrix is itself not a martingale. Of course, for special situations where 
$[M_{\alpha_I},\theta_n]\,=\,0$ for every $(I,n)$, the $\theta$'s are martingales but we are seeking the answers for generic $\theta_0$ 
and subseqently, generic $\theta_n$'s.

We next look at the matrix elements of $\theta$'s in the $|q_i\rangle$ basis. Introduce $\theta_n^{ij}\,=\,\langle\,q_i|\theta_n|q_j\rangle$. 
Let us see under what conditions, if at all, these are martingales:
\bea
\label{eq:rhomemartingale}
\mathbb{E}[\theta_{n+1}^{ij}|\theta_n\ldots,\,\theta_0]\,&=&\,\sum_{\alpha_{i_{n+1}}}\,P(\alpha_{i_{n+1}}|\alpha_{i_1},\ldots,\alpha_{i_n})\,\frac{\langle\,q_i|M_{\alpha_{i_{n+1}}}\,\theta_n\,M_{\alpha_{i_{n+1}}}^\dag|q_j\rangle}{Tr\,M_{\alpha_{i_{n+1}}}\,\theta_n\,M_{\alpha_{i_{n+1}}}^\dag}\nonumber\\
&=&\,\sum_{\alpha_{i_{n+1}}}\,\lambda_{\alpha_{i_{n+1}}}^i{\lambda_{\alpha_{i_{n+1}}}^j}^*\,\langle\,q_i|\theta_n|q_j\rangle\nonumber\\
&=&\,\sum_{\alpha_{i_{n+1}}}\,\lambda_{\alpha_{i_{n+1}}}^i{\lambda_{\alpha_{i_{n+1}}}^j}^*\,\theta_n^{ij}\,=\,{\tilde\mu}_{ij}\,\theta_n^{ij}
\eea
where
\be
\label{eq:mutilde}
{\tilde\mu}_{ij}\,=\,\sum_{\alpha_{i_{n+1}}}\,\lambda_{\alpha_{i_{n+1}}}^i\,{\lambda_{\alpha_{i_{n+1}}}^j}^*
\ee
As ${\tilde\mu}_{ij}$ is complex, nothing can be said about the martingale property of $\theta_n^{ij}$. As $|{\tilde\mu}_{ij}|$ is just 
the $\mu_{ij}$ introduced earlier, it is suggestive to look at the martingale properties of $A_n^{ij}\,=\,|\theta_n^{ij}|$ instead. Indeed
\be
\label{eq:Amartingale}
\mathbb{E}[A_{n+1}^{ij}|\theta_n,\ldots,\theta_1]\,=\,\mu_{ij}\,A_n^{ij}
\ee
Depending on whether $\mu_{ij}$ is 1 or less than 1, $A_n^{ij}$ is a martingale or a supermartingale. As already discussed, the diagonals
$\mu_{ii}\,=\,1$ for every i. Consequently, $A_n^{ii}$ are martingales. When $i\ne\,j$, $\mu_{ij}$ can be one or less than one, depending on
whether $|q_i\rangle$ and $|q_j\rangle$ are degenerate or not. If they are, $\mu_{ij}=1$ and the corresponding $A_n^{ij}$ are martingales.
Else, they are supermartingales.

Note that as the density matrices $\theta_n$ are hermitian, the diagonal elements $\theta_n^{ii}$ are real and hence in $A_n^{ii}$ the
modulus is redundant i.e. $A_n^{ii}\,=\,\theta_n^{ii}$. These can also be alternately expressed as $Tr\,\theta_n\Pi_i$ with $\Pi_i$ being
the one-dimensional projective operator $|q_i\rangle\langle\,q_i|$ introduced before.

This suggests testing $A_n^{{\bar{i}}}\,=\,\theta_n^{{\bar{i}}}\,=\,Tr\,\theta_n\,\Pi_{{\bar{i}}}$ where $\Pi_{{\bar{i}}}$ are the 
$d_{{\bar{i}}}$-dimensional projection operators onto the degenerate subspaces labelled by ${{\bar{i}}}$. On using eqn.(\ref{eq:degproj}), it
is easily seen that $\theta_n^{{\bar{i}}}$ are also real, and hence their equality with $A_n^{{\bar{i}}}$.

We now present the martingale test:
\bea
\label{eq:thetadegmart}
\mathbb{E}[\theta_{n+1}^{{\bar{i}}}|\theta_1,\ldots,\theta_n]\,&=&\,\sum_{\alpha_{i_{n+1}}}\,\frac{Tr\,(M_{\alpha_{i_{n+1}}}\,\theta_n\,M_{\alpha_{i_{n+1}}}^\dag\,\Pi_{{\bar{i}}})}{P(\alpha_{i_{n+1}}|\alpha_{i_1},\ldots,\alpha_{i_n})}\cdot\,P(\alpha_{i_{n+1}}|\alpha_{i_1},\ldots,\alpha_{i_n})\nonumber\\
\,&=&\,\sum_{\alpha_{i_{n+1}}}\,|\lambda_{\alpha_{i_{n+1}}}^{{\bar{i}}}|^2\,Tr\,\theta_n^{{\bar{i}}}\Pi_{{\bar{i}}}\nonumber\\
\,&=&\,Tr\,\theta_n\Pi_{{\bar{i}}}
\eea
Thus proving that $\theta_n^{{\bar{i}}}$ is indeed a martingale. In going to the second step, we invoked eqn.(\ref{eq:projeigendeg}) twice
in conjunction with cyclicity of traces, and, in going to the last step we used eqn.(\ref{eq:lambpropdeg1}).

\subsection{Boundedness and Asymptotic Convergence}
We know show that the quantities considered above with martingale and submartingale properties are also bounded. First let us consider
the diagonal matrix elements $\theta_n^{ii}$. As $\theta_n$ are density matrices, they satisfy
\be
\label{eq:trrho}
\sum_i\,\theta_n^{ii}\,=\,1
\ee
That $\theta_n$ are density matrices, which are positive-semi definite, means that 
\be
\label{eq:rhopos}
\theta_n^{ii}\,\ge\,0
\ee
These two equations put together imply that $\theta_n^{ii}$ are bounded for every $i$.

To show that $A^{ij}\,=\,|\theta^{ij}|$ are bounded for any state $\theta$, consider
\be
\label{eq:offdiagbound}
Tr\,\theta^2\,=\,\sum_{ij}\,\theta^{ij}\theta{ji}\,=\,\sum_{ij}\,|\theta^{ij}|^2
\ee
where we have used the fact that density matrices are also hermitian. But $Tr\,\theta^2$, also called purity, is also positive - definite
and bound by 1. Therefore each of $|\theta^{ij}|$ is also bounded.

Finally, we show that $\theta^{{\bar{i}}}\,=\,Tr\, \theta\Pi_{{\bar{i}}}$ are also positive semi-definite and bounded. It follows
straightforwardly from eqn.(\ref{eq:degproj}) with $i_0$ labelling some orthonormal basis spanning the degenerate subspace ${{\bar{i}}}$.
Then it follows that
\be
\label{eq:degprojpos}
Tr\,\theta\Pi_{{\bar{i}}}\,=\,\sum_{i_0}\,Tr\,\theta\Pi_{{\bar{i}},i_0}
\ee
Since each term is positive semi-definite and bounded by unity, so the sum is also positive semi-definite and bounded by $d_{{\bar{i}}}$. But
the sum also satisfies
\be
\label{eq:degprojbound}
Tr\,\theta\Pi_{\bar{i}}\,\le\,Tr\,\theta\,=\,1
\ee
It is actually bounded by unity too.

Consequently, the martingale convergence theorem applies to each one of them. In other words, $A_n^{ij}\rightarrow\,A^{ij,\infty}$ as
$n\rightarrow\,\infty$, and, $\theta_n^{{\bar{i}}}\rightarrow\,\theta^{{\bar{i}},\infty}$ in the same limit. For the diagonal case, $\theta_n^{ii}\rightarrow\,\theta^{ii,\infty}$.

It should be emphasized that at this stage one can not claim $\theta_n\rightarrow\,\theta_\infty$ as $n\rightarrow\,\infty$ for the
entire density matrix, but only for its diagonal elements and the magnitude of the off-diagonal matrix elements. That is why we avoided
the notations $\theta_\infty^{ii},A_\infty^{ij}$ for the asymptotic limits. This is a subtle but important point.

The asymptotic form of the martingale condition of eqn.(\ref{eq:Amartingale}) takes the form
\be
\label{eq:Amartasymp}
A^{ij,\infty}\,=\,\mu_{ij}\,A^{ij,\infty}
\ee
It is easy to see that  the map
\be
\label{eq:asympfix}
\theta_{n+1}\,=\,\frac{M_{\alpha_{i_{n+1}}}\,\theta_n\,M_{\alpha_{i_{n+1}}}^\dag}{Tr\,M_{\alpha_{i_{n+1}}}\,\theta_n\,M_{\alpha_{i_{n+1}}}^\dag}
\ee 
leads to
\be
\label{eq:Amap}
A_{n+1}^{ij}\,=\,\frac{|\lambda_{\alpha_{i_{n+1}}}^i||\lambda_{\alpha_{i_{n+1}}}^j|\,A_n^{ij}}{\sum_k\,|\lambda_{\alpha_{i_{n+1}}}^k|^2\,A_n^{kk}}
\ee
which in turn gives to the asymptotic relation
\be
\label{eq:Amapasymp}
A^{ij,\infty}\,\sum_k\,|\lambda_{\alpha_I}^k|^2\,\theta^{kk,\infty}\,=\,|\lambda_{\alpha_I}^i||\lambda_{\alpha_I}^j|\,A^{ij,\infty}
\ee
for all $(i,j,I)$. There is a subtelety in arriving at this. While the various $A_n^{ij}$ do take their asymptotic forms, the asymptotic
behaviour of $\alpha_{i_{n+1}}$ is more subtle. There is nothing that demands that these take a limiting value and hence for very large values 
of n, these can take all possible values randomly and is denoted by $\alpha_I$. To appreciate this better, consider the initial state of the
system to be in an eigenstate $|q_M\rangle$. Subsequently, though the system state remains the same by virtue of the QND nature of the
measurements, the probe outcomes need not be. They are randomly generated by the probability distribution 
$P(\alpha_{i_n})\,=\,|\lambda_{\alpha_{i_n}}^M|^2$. 

We now work out the detailed consequences of these asymptotic relations. Once again, it is worth emphasizing that only limiting forms of
$A_n^{ij}$ have appeared explicitly and not the limiting form of the entire density matrix. In fact, we still do not know whether the entire 
density matrix itself takes any limiting form.
\section{Consequences}
\label{sec:consequences}
From eqn.(\ref{eq:Amartasymp}) it is seen that the diagonals $A^{ii,\infty}$ are unrestricted at this stage because of $\mu_{ii}\,=\,1$.
\subsection{Purely non-degenerate case}  
\label{subsec:nondegasymp}
Let us first consider the case when ${\hat q}$ has a purely non-degenerate spectrum. It follows that the measurement operators $M_{\alpha_I}$
also have purely non-degenerate spectrum for every I. Now $\mu_{ij}\,\le\,1$ for $i\ne\,j$. Eqn.(\ref{eq:Amartasymp}) immediately has the
consequence
\be
\label{eq:ndegoffdiagasymp}
A^{ij,\infty}\,=\,0 \quad\quad i\ne\,j
\ee
Thus asymptotically only diagonal entries are non-vanishing.

This raises the question as to how many diagonal elements can be non-vanishing at the same time? Suppose that both $A^{ii,\infty}$ and 
$A^{jj,\infty}$ are 
non-vanishing. On using eqn.(\ref{eq:Amapasymp}) for diagonals,
\be
\label{Amapasympdia}
A^{ii,\infty}\,\sum_k\,|\lambda_{\alpha_I}^k|^2\,\theta^{kk,\infty}\,=\,|\lambda_{\alpha_I}^i|^2\,A^{ii,\infty}
\ee
Hence, $A^{ii,\infty}\,\ne\,0$ implies
\be
\label{eq:Aiiasympne0}
|\lambda_I^i|^2\,=\,\sum_k\,|\lambda_I^k|^2\,A^{kk,\infty}
\ee
The r.h.s is actually independent of $i$. Likewise, $ A^{jj,\infty}\,\ne\,0$ implies
\be
\label{eq:Ajjasympne0}
|\lambda_I^j|^2\,=\,\sum_k\,|\lambda_I^k|^2\,A^{kk,\infty}
\ee
Combining the two, one gets
\be
\label{eq:2diane0}
|\lambda_I^i|^2\,=\,|\lambda_I^j|^2\quad\quad \forall I
\ee
But as already shown, this is only possible if $|i\rangle,|j\rangle$ are {degenerate}, contradicting the purely non-degenerate spectrum
of ${\hat q}$.

Therefore we reach the very important conclusion that when ${\hat q}$ has only non-degenerate spectrum, only one of the diagonals can be 
non-vanishing. Let it happen for, say, $i\,=\,N$.Combining with the fact that all off-diagonals vanish, one can conclude that the 
density matrix itself approaches the asymptotic value
\be
\label{eq:asymptheta}
\theta_\infty\,=\,|q_N\rangle\langle\,q_N|
\ee
In other words, this particular trajectory asymptotically converges to the observable eigenstate $|q_N\rangle$. Equivalently, trajectories can
be labelled by the eigenvalues of the observables. There will be family of trajectories all labelled by the same index. Different families 
will be labelled by different eigenstates. This is exactly the picture that was obtained in \cite{weakrepeat2014}. We will address the
the probabilities with which different families occur shortly.
\subsection{Degenerate observables and L\"uders rule}
\label{subsec:luders}
Now we turn to situations where ${\hat q}$ also has degeneracies in it's spectrum. Let us say that $|q_i\rangle$ and $|q_j\rangle$ are now
degenerate. As discussed in detail before, $\lambda_{\alpha_I}^i\,=\,\lambda_{\alpha_I}^j$, for every I. From the previous subsection, one
immediately sees that both $A_n^{ii}$ and $A_n^{jj}$ can be nonvanishing.

Additionally, since $\mu_{ij}\,=\,1$ even though $i\ne\,j$, there is no reason for the off-diagonals $A_n^{ij}$ to vanish. Obviously, this can 
be extended to all states in the same degeneracy class and the asymptotic state is a $d_{{\bar{i}}}\times\,d_{{\bar{i}}}$ density matrix,
spanned by the degenerate subspace. 
This, however, does not imply that asymptotically the state itself reaches a limiting value, as the matrix elements within this block can
still evolve. The martingale and the martingale convergence in this case are not enough to ensure convergence of the state itself. For that,
one has to use eqns.(\ref{eq:degproj},\ref{eq:projeigendeg}) and eqn.({\ref{eq:degproj}). The first of them says that the projection operators
$\Pi_{\bar{i}}$ act as identity on thesubspace ${\bar{i}}$, and the second  that the measurement
are essentially identity operators(actually $(\lambda_{\alpha_I}^{{\bar{i}}}{\bf I})$ while acting on the 
degenerate subspace. It should be noted that the QND nature of measurements is crucial for this. This means whatever was the projection of 
the initial unknown state $\theta_0$ onto the degenerate subspace, it remains
the {\bf same} throughtout the trajectory. 
In particular, the asymptotic state for this family of trajectories will also be this projection. 
This is stated explicitly in eqn.(\ref{eq:asythetabar}) in the next subsection.
This is exactly the L\"uders prescription.

{\bf Not all states can purify!}: Consider initial states such that their projection onto a particular degenerate subspace are mixed states.
Then as per our reasoning presented, throughout the trajectory the projection onto that degenerate subspace remains the same mixed state. In
particular the asymptotic states can only be the same mixed states. 
Note that
Our result does not contradict that of Maassen and K\"ummerer \cite{maassen2006}, who also used a discrete-time approach. In their
analysis, under certain conditions, the system may converge to a subspace, referred to as the dark subspace, rather
than to a pure state. Here, we show that a QND measurement with a degenerate spectrum falls into this category,
with the dark subspace being precisely one of the degenerate subspaces.
\subsection{An improved treatment of degeneracies}
\label{subsec:ludersimproved}
We now present an improved and elegant treatment of the situation with degenerate observables based on projections onto degenerate subspaces. 
Using
eqn.(\ref{eq:projeigendeg}) and eqn.(\ref{eq:degprojprop}), it is easy to work out the measurement maps in the form
\be
\label{eq:ibarmap}
\theta_{n+1}^{{\bar{i}}}\,=\,\frac{|\lambda_{\alpha_{n+1}}^{{\bar{i}}}|^2\,\theta_n^{{\bar{i}}}}{\sum_{{\bar{k}}}\,|\lambda_{\alpha_{i_{n+1}}}^{{\bar{k}}}|^2\,\theta_n^{{\bar{k}}}}
\ee
This, as before, leads to the important relation for the asymptotic $\theta^{{\bar{i}},\infty}$:
\be
\label{eq:ibarasymp}
\theta^{{\bar{i},\infty}}\,\sum_{{\bar{k}}}\,|\lambda_{\alpha_I}^{{\bar{k}}}|^2\,\theta^{{\bar{k}},\infty}\,=\,|\lambda_{\alpha_I}^{{\bar{i}}}|^2\,\theta^{{\bar{i}},\infty}
\ee
where once again we have introduced a generic $\alpha_I$ for the asymptotic sequence of probe outcomes, and have avoided the explicit 
introduction of any asymptotic density matrix. From this asymptotic version of the map, the condition for non-vanishing 
$\theta^{{\bar{i}},\infty}$ is:
\be
\label{eq:ibarasyne0}
|\lambda_{\alpha_I}^{{\bar{i}}}|^2\,=\,\sum_{{\bar{k}}}\,|\lambda_{\alpha_I}^{{\bar{k}}}|^2\,\theta^{{\bar{k}},\infty}\quad\quad \forall I
\ee
As before, we ask whether $\theta^{{\bar{i}},\infty}\,\ne\,0$ for more than one $i$ simultaneously. Let us analyse this for two distinct 
values of $i$, say, 
$i_1,i_2$. Since r.h.s of eqn.(\ref{eq:ibarasyne0}) is independent of $i$, one obtains:
\be
\label{eq:2ibarne0}
|\lambda_{\alpha_I}^{{\bar{i}}_1}|^2\,=\,|\lambda_{\alpha_I}^{{\bar{i}}_2}|^2\quad\quad \forall I
\ee
As per the properties of $\lambda$'s derived earlier, this can only happen if states in the degeneracy classes ${\bar{i}}_1$ and ${\bar{i}}_2$
are also degenerate. That can not happen as by construction ${\bar{i}}$ label distinct degeneracy classes. Hence eqn.(\ref{eq:ibarasyne0})
can only be satisfied for only one value of ${\bar{i}}$, say, ${\bar{M}}$. 

As elaborated in the previous subsection, this alone is insufficient to argue that the state of the system itself converges asymptotically
along trajectories. The properties of the QND measurements are crucial for this conclusion. Again, eqns.(\ref{eq:degproj},\ref{eq:projeigendeg})
are needed to show that whatever was the projection of the initial state onto the degenerate subspace, continues to be the projection onto
the degenerate subspace at every step of the trajectory, and hence asymptotically too.
Then, the asymptotic state along this trajectory will be
\be
\label{eq:asythetabar}
\theta_\infty\,=\, \frac{\Pi_{{\bar{M}}}\theta_0\Pi_{{\bar{M}}}}{Tr\,\Pi_{{\bar{M}}}\theta_0}
\ee
Clearly only those ${\bar{i}}$ can be possible asymptotic states for which $Tr\,\Pi_{{\bar{i}}}\,\theta_0\,\ne\,0$. This is indeed the
L\"uders rule.

The advantages of using distinct degeneracy classes should now be obvious. In terms of them, the description is essentially of the same
mathematical form as what was encountered for the case with purely non-degenerate spectrum albeit describing situations with degeneracy.
The non-degenerate part of the spectrum is taken care of by the same formalism by those ${\bar{i}}$ for which the dimensionalities 
$d_{{\bar{i}}}$ are 1.

\subsection{Born rule for trajectories}
Though we have now established that each trajectory is labelled by its asymptotic state ${\bar{M}}$, the question still remains as to
the probability distribution $P({\bar{M}})$  for the trajectories.The 2014 analysis \cite{weakrepeat2014} had strongly hinted it would be 
$\theta_0^{{\bar{M}}}\,=\,Tr \theta_0\,\Pi_{{\bar{M}}}|$

That was based on the reasonable, yet heuristic, expectation that ensemble averages of repeated measurements can be thought of as a two-step
averaging, the first along a given trajectory, and the final one over all possible trajectories. Having learnt how to handle individual
trajectories, we shall now put that heuristic expectation 
on a firmer basis.

What entered the martingale results were {\it Conditional Expectations} $\mathbb{E}[x_{n+1}|x_n,\ldots,x_1]$ whereas ensemble averages are 
{\it Unconditional Expectations} $\mathbb{E}[x_{n+1}]$. In what follows we shall base our considerations on the martingales 
$\theta_n^{{\bar{i}}}$. Submartingales and supermartingales, if at all useful, may require more sophisticated approaches. So what we need are 
$\mathbb{E}[\theta_n^{{\bar{i}}}]$. We give two proofs to bring out different aspects of the problem.
The conditional and unconditional expectations are related by:
\bea
\label{eq:unconditionalE}
\mathbb{E}[\theta_{n+1}^{{\bar{i}}}]\,&=&\,\sum_{\alpha_{i_n},\ldots,\alpha_{i_1}}\,\mathbb{E}[\theta_{n+1^{{\bar{i}}}}|\theta_n^{{\bar{i}}}\ldots\,\theta_1^{{\bar{i}}}]P(\alpha_{i_n}\ldots\,\alpha_{i_1})\nonumber\\
&=&\,\sum_{\alpha_{i_n},\ldots,\alpha_{i_1}}\,\theta_n^{{\bar{i}}}\,P(\alpha_{i_1},\ldots,\alpha_{i_n}\nonumber\\
&=&\,\sum_{\alpha_{i_n},\ldots,\alpha_{i_1}}\,\frac{|\lambda_{\alpha_{i_n}}^{{\bar{i}}}|^2\,\theta_{n-1}^{{\bar{i}}}}{P(\alpha_{i_n}|\alpha_{i_1},\ldots,\alpha_{i_{n-1}})}\cdot\,P(\alpha_{i_n}|\alpha_{i_n},\ldots,\alpha_{i_1})P(\alpha_{i_1},\ldots,\alpha_{i_{n-1}})\nonumber\\
&=&\,\sum_{\alpha_{i_{n-1}},\ldots,\alpha_{i_1}}\,\theta_{n-1}^{{\bar{i}}}\,P(\alpha_{i_1},\ldots\,\alpha_{i_{n-1}})\nonumber\\
&=&\,\mathbb{E}[\theta_{n-1}^{{\bar{i}}}]
\eea
Repeating this, one arrives at
\be
\label{eq:uncondall}
\mathbb{E}[\theta_{n+1}^{{\bar{i}}}]\,=\,\mathbb{E}[\theta_n^{{\bar{i}}}]\,=\,\ldots\,=\,\theta_0^{{\bar{i}}}
\ee
This implies that in the limit $n\rightarrow\infty$,
\be
\label{eq:uncondasymp}
\mathbb{E}[\theta^{{\bar{i}},\infty}]\,=\,\theta_0^{{\bar{i}}}
\ee 
We now present an alternative way of evaluating $\mathbb{E}[\theta_{n+1}^{{\bar{i}}}]$. Instead of removing all conditionalities in one go
as above, we shall remove the conditionalities one by one. We explicity show how to remove the conditionality on $\theta_n^{\bar{i}}$:
\bea
\label{eq:uncondstep}
\mathbb{E}[\theta_{n+1}^{{\bar{i}}}|\theta_{n-1}^{\bar{i}},\dots,\theta_0^{\bar{i}}]\,&=&\,\sum_{\alpha_{i_n}}\,\mathbb{E}[\theta_{n+1}^{\bar{i}},\ldots,\theta_0^{\bar{i}}]\,P(\alpha_{i_n}|\alpha_{i_{n-1}},\ldots,\alpha_{i_1})\nonumber\\
&=&\,\sum_{\alpha_{i_n}}\,\theta_n^{\bar{i}}\,P(\alpha_{i_n}|\alpha_{i_{n-1}},\ldots,\alpha_{i_1})\nonumber\\
&=&\,\sum_{\alpha_{i_n}}\,\frac{|\lambda_{\alpha_{i_n}}^{\bar{i}}|^2\,\theta_{n-1}^{\bar{i}}}{P(\alpha_{i_n}|\alpha_{i_{n-1}},\ldots,\alpha_{i_1})}\cdot\,P(\alpha_{i_n}|\alpha_{i_{n-1}},\ldots,\alpha_{i_1})\nonumber\\
&=&\,\theta_{n-1}^{\bar{i}}
\eea
In going to the second line we used the martingale condition, the map in going to the third, and the properties of $\lambda$'s in arriving
at the last. Repeating the step to remove all conditionalities one arrives at
\be
\label{eq:uncondall2}
\mathbb{E}[\theta_{n+1}^{\bar{i}}]\,=\,\theta_0^{\bar{i}}
\ee
In the asymptotic limit one again gets back eqn.(\ref{eq:uncondasymp}).

Now, $\theta^{{\bar{i}},\infty}$ is a random variable, taking the value 1 if the asymptotic states are on trajectories labelled by
${\bar{i}}$, and 0 otherwise (recall that this is because only one degeneracy class survives asymptotically). On the other hand, the unconditional expectation $\mathbb{E}[\theta^{{\bar{i}},\infty}]$ is given by:
\be
\label{eq:trajprob}
\mathbb{E}[\theta^{{\bar{i}},\infty}]\,=\,P({\bar{i}})
\ee
Combining this with eqn.(\ref{eq:uncondasymp}), and recalling that $\theta_0^{\bar{i}}\,=\,|c_{{\bar{i}}}|^2$, the probability of finding the
degenerate subspace ${\bar{i}}$ in $\theta_0$, one gets
\be
\label{eq:bornibar}
P({\bar{i}})\,=\,|c_{{\bar{i}}}|^2
\ee
This is just the Born rule, generalized to include the L\"uders rule.

Thus one concludes that repeated measurements on single copies of quantum systems, giving rise to trajectories, are surprisingly similar
to projective measurements. Each trajectory is akin to a particular outcome of projective measurements, which are eigenvalues of the 
observable and random. The system state becomes the corresponding eigenstate. Thus on a single copy, one of the many possible trajectories 
are randomly realised and the asymptotic state along that trajectory is one of the eigenstates (degenerate or otherwise). This can not
determine the unknown initial state of the system, very much like the situation in projective, also called strong, measurements.
\section{Comparisons with other works}
Now we make a comparison of our work with others that addressed the same issue. The first of these is the work of Alter and Yamomoto 
\cite{alterbook2001,alter1,alter2}. After finishing our work in 2019, we became aware of the works of Bauer et al 
\cite{bauershort2011,bauerlong2013}, as well as by Amini et al \cite{amini2011}. Before making a detailed comparison of our work with
theirs, we turn to even earlier works by Alter and Yamamoto.
\subsection{Early works of Alter and Yamomoto}
The earliest claims of the impossibility of determining the unknown initial state of a single copy are due to Orly Alter and Yoshihisa Yamamoto
\cite{alterbook2001,alter1,alter2}. Alter's book \cite{alterbook2001} gives a detailed account of their arguments. Their essential argument
is that the `statistics' of such measurements can not, even in principle, lead to a determination of the initial state.

As this is also a manifestation of our results, we have tried hard to relate their approaches and conclusions to ours. This has not been
easy as both our approach as well as those of Bauer et al \cite{bauershort2011,bauerlong2013},and, of Amini et al \cite{amini2011} are
more less direct consequences of quantum theory, the Alter-Yamomoto approaches seem to hinge on what they call `estimation'. While probability
distributions can be determined by ensemble measurements, and they call the peak of such probability distributions as the estimated value, in
measurements on single copies the probability distribution can not be measured and no such estimation can be meaningfully carried out. In
their sec.(2.3) they seem to imply that the single outcome of the probe measurement can in itself be used for such an estimation.

We don't see how that can be and even if one goes by that, why such an estimate has any significance at all. Such a recipe would also
give zero variance, which the authors are also aware of. There are notational confusions($q$ is denoted for system variables while ${\tilde q}$ 
used for probe variables), yet they seem to get mixed up with introduction of $\delta(q\,-\,{\tilde q}_1)$ etc. Going into more technical 
details, their expression $P({\tilde q_2}|{\tilde q}_1)$ for the conditional probability for obtaining second probe outcome given the first
probe outcome (eqn.(2.8) of \cite{alterbook2001} does not at all agree with our expressions for conditional probabilities of 
eqn.(\ref{eq:condprob2|1p})). In fact their expression does not even involve the second measurement operator.

Even though they too found widths of photon distributions to go to zero asymptotically in their example of photon number measurements
\cite{alter1}, indicating that trajectories end in eigenstates asymptotically, that they do randomly on possible eigenstates with probabilities
given by Born rule is not very transparent. Their phoiton number measurement example is a special case of our sec.(\ref{sec:gaussianpovm}).
Nevertheless, with the final claims
being related, it is imperative to arrive at a closer dictionary between our approaches.
\subsection{Works of Bauer et al and Amini et al}
We first list the common aspects of of our works with these:
\begin{itemize}
\item All are based on the same formalism for generalized measurements.
\item All make use of measurement operators and their properties except \cite{bauershort2011}. But in their longer work \cite{bauerlong2013},
they too use measurement operators explicitly.
\item All are based on the same characterization of QND measurements. Even though this is not explicit in \cite{bauershort2011}, it is
made explicit in \cite{bauerlong2013}
\item Except for \cite{bauershort2011} which only treats pure initial states, all three treat both pure and mixed initial states.
\item All three make use of Martingales. 
\item The proof of Born rule for trajectories is the same in all three cases.
\item None of the works makes use of continuous time measurements and stochastic differential equations, and all derivations follow directly 
from quantum mechanics.
\end{itemize}
In what follows we shall compare our work only with \cite{bauerlong2013}.
\subsection{Works of Amini et al}
In \cite{amini2011} they only treat the diagonal $A_n^{ii}$ and prove their martingale property. Since they only consider diagonals, the 
modulus becomes unnecessary as the diagonal elements of the hermitian density matrices are always real. They use the same form of the QND 
condition as ours i.e. the measurement operators are diagonal in the basis spanned by the eigenstates of the system observable. This is
stated as their assumption I.

Since they do not analyse the off-diagonal $A_n^{ij}$, they are unable to handle the case of system observables with degenerate spectra.
They also make a further assumption(their assumption II) which is completely equivalent to our requirement that for non-degenerate
states $|\lambda_{\alpha_I}^i|^2\,\ne\,|\lambda_{\alpha_I}^j|^2$ for at least one $I$.

They do not apply the martingale convergence theorem. Their proofs of convergence are rather cumbersome. They consider convex functions
of the density matrix and show that they are submartingales. After some tedious algebra they arrive at the result that asymptotically
the state on a given trajectory approaches an eigenstate.

Our proofs are considerably simpler and straight-forward.
\subsection{Work of Bauer et al}
As already mentioned we will only compare \cite{bauerlong2013} with our work as it is much closer in technical details to us than 
\cite{bauershort2011}. A word of caution right at the outset is that Bauer et al call the system eigenstates as {\it Pointer states}
whereas we have used that terminology for probe states. 

They do not state explicitly the relation between the measurement interaction represented by a unitary transformation U,and the system
observable measured. In our case we explicitly require U to be depend only on the system observable. This lack of explicitness makes
their discussion needlessly obscure. Whereas in our case the criterion for measurements to be QND is that $[U,{\hat q}]=0$ manifesting
expplicitly in the QND condition that the system eigenstates are also eigenstates of measurement operators, this is not evident in 
\cite{bauerlong2013}. However, they do state that their QND criterion is that their measurements preserve a preferred basis of the system
Hilbert space ${\cal H}_S$ (para before their eqn.(2)). But the relationship of that preserved basis to the eigenstates of the system 
observable is not at all clear.

While our specification of U and our QND criterion is compatible with theirs, it is far from clear whether their general characterizations admit
other possibilities. This requires addressing the rather difficult issues of {\it Fixed Points of Quantum Maps} in all their generality. See
\cite{quopfixed2002} for a more rigorous approach. 

For the same reasons, it is also not easy to see whether their treatment includes degenerate system observables are not. Without reference
to the measured observable, they first consider an assumption to the effect that for every pair of system states $(\alpha,\beta)$ there
exists at least one probe outcome for which the probabilities of outcome are not the same ( see the opening para of their section(4.)) which
is equivalent to our properties of $\lambda$'s for the case of fully non-degenerate system observables. In their sec.(5.) \cite{bauerlong2013}
relax this condition.

It is noteworthy to see how they handle the cases with degeneracies. They do not consider the martingale properties of off-diagonal entries
which was essential to our first treatment of the degenerate cases. They instead use the notion of {\it sectors} which they introduce based
on their notion of equivalence classes among system states wherein two states are equivalent if their measurement statistics are identical.
This is completely equivalent to our use of degeneracy classes and projections onto them. We too did not have to consider the martingale
properties of off-diagonal entries this way. Actually we were inspired by \cite{bauerlong2013} in this.

Though \cite{bauerlong2013} also derive martingale properties, and make use of the martingale convergence theorem, their approach to 
asymptotic convergence is still rather formal based heavily on classical probability theory and measure theory. 
Although Bauer et al. provide a comprehensive measure-theoretic treatment of
martingales, the abstraction level can make the mechanism behind convergence less accessible. We therefore present
a more explicit construction tailored to QND measurements.
We leave it to the reader to make a more detailed comparison. In contrast, our approach relies on a simple and direct use of 
the martingale convergence theorem along with the use of very simple and transparent properties of the eigenvalues of the measurement 
operators to arrive at the desired asymptotic behaviours. 

A major difference between our approach and theirs consists in the following: while in our schema for repeated measurements, the measurements
are identical at each step, \cite{bauerlong2013} allow this to vary at each step. More precisely, they characterize a measurement setup
by the triplet $(|\Psi\rangle_P, U, |i\rangle_P)$ where the first element is the initial probe state, U the measurement interaction and the
last element is the orthonormal basis for the probe Hilbert space ${\cal H}_P$. They allow for this triplet to change arbitrarily along
the trajectory. This is an enormous flexibility, potentiall very useful in actual realisations.

It is worth understanding the basis for this flexibility. That lies in the fact that the martingale conditions are {\it local} along
the trajectory in the sense that the condition at the nth step does not actually depend on how the previous states were realized but only
on their values. For this reason, our analysis can also be trivially extended to have this flexibility. We shall make use of this flexibility
to show how our results remain robust even after one takes into account the inevitable free evolutions of both the system and the probe
in between the generalized measurements. Otherwise, the results of the asymptotic behaviours would have been only of academic interest. 

\section{Free Evolution}
\label{sec:freeevol}
In all the three approaches so far the effects of the free evolutions of the system as well as the probe in between two measurements were
not taken into account. For the results so far obtained to have any practical relevance, these inevitable evolutions must be factored in.
Denoting the free hamiltonians of the probe and system by $H_P,H_S$, and the time in between nth and the next measurement by $\tau_n$(this 
need not be uniform in n), the unitary transformation resulting from the free evolutions is
\be
\label{eq:freeU}
U_{free}\,=\,e^{iH_P\tau_n}\cdot\,e^{iH_S\tau_n}
\ee 
It is clear that if $U_{n+1}$ was the measurement interaction at the $(n+1)$-th step in the absence of the free evolutions, taking them
into account would effectively change the measurement interaction according to
\be
\label{eq:freeevol}
U_{n+1}\rightarrow\,U_{n+1}^\prime\,=\,U_{n+1}U_{free}
\ee
And by the last mentioned flexibility in choosing the measurement interaction arbitrarily at each stage, it would appear that the
conclusions about the asymptotic behaviour of trajectories would be robust against the effect of free evolutions.

However, there is a catch and that has to do with the fact that whatever measurement interactions are chosen, they must remain of the
QND type. While the free evolution of the probe does not affect the QND nature, that is certainly not true of any free evolution of the
system.

Remarkably, another feature of QND measurements that was not invooked till now comes to the rescue. This is that the system observable
${\hat{q}}$ must be a constant of motion under the total Hamiltonian $H\,=\,H_P+H_S+H_I$(see eqn.4.34 of \cite{brakha1992}). This in turn
implies
\be
\label{eq:freeqnd}
[H_S,{\hat{q}}]\,=\,0
\ee
which in turn implies that the QND condition is not affected by free evolutions.
\section{Anti-Zeno effect?}
\label{sec:againstzeno}
The quantum Zeno paradox(effect) first introduced by Misra and Sudarshan \cite{qzeno1977} has generated tremendous interest, and literature.
Originally formulated for ensembles and based on continuous measurements, it essentially asserted that a continually observed quantum
system does not involve, leading to the folklore 'a watched pot doesn't boil over'. The corresponding effect in single copies is a lot
more subtle and has been addressed in chapter 6 of \cite{alterbook2001}. We point out that the results of our work, and of amini and Bauer,
can be interpreted as a counter to the quantum zeno effect. This is in the sense that asymptotically, repeated measurements not only do not
stop the system from evolving, they actually drive it unpredictably away from where they started. In that sense, we have a situation where
a continually(but not continuously) watched pot not only boils over, but does so unpredictably. This should not be confused with the
{\it inverse zeno effect} discussed by Aharonov and Vardi \cite{invqzeno}.
\section{Conclusions}
In this paper we have thoroughly investigated asymptotic behaviour of quantum trajectories generated by POVM QND measurements. 
We have used
the discrete time approach as against methods based on continuous measurements and stochastic differential equations(SDE). The main results
obtained are:
\begin{itemize}
\item Asymptotically, in the sense of after very large number of repeated measurements, the state of the system approaches one of the
eigenstates of the system.
\item When the system observable is degenerate, the asymptotic state is also in one of the degenerate subspaces. In particular, 
it exactly equals the projection of the initial system state onto the particular subspace, thus reproducing
L\"uders rule.
\item Our analysis is very general and applies to initial system states being pure or mixed, and to system observables with or without
degeneracies.
\item It is also proved that the probability distribution for the quantum trajectories, now labelled by the asymptotic states, is given by
the Born rule.
\item Though the initial objective was to seek results for weak measurements, the results, surprisingly, hold for arbitrary generalized
measurements(POVM) as long as they are of the QND type.
\item This may be understood as follows: one can associate a time scale with measurements. Projective or strong measurements with short
time scales and weak measurements with very long, but finite, time scales. In fact the von Neumann model for projective measurements can
be realized through {\it impulsive}(very short-time) measurements. To reach asymptotia in repeated measurements, the total times have to be so large that the asymptotic behaviour is insensitive to whether the measurements are strong or weak. It would be very important to experimentally
establish this.
\item This feature was already present in our earlier work \cite{weakrepeat2014} but in our early emphasis on weak measurements, we had overlooked this.
\item The results were established via the powerful concepts of martingales and martingale convergence theorems. Our proofs are particularly simple
and straightforward, without invoking highly mathematical aspects of probability theory. We have established a number of important properties
of generalized POVM's which facilitated these proofs.
\item A striking aspect highlighted by \cite{bauerlong2013} was that the asymptotic behaviours are robust against changing the measurement
schemes at each step of the random walk. This feature should be investigated in more detail both experimentally and theoretically.
\item We have used this in showing that free evolutions of both the probe and the system do not affect the asymptotic behaviours.
\end{itemize}

A number of beautiful experiments have already demonstrated many key features of the asymptotic behaviours discussed here. In fact, it was
the experiments of Vijayaraghavan and collaborators \cite{vijayexpt2016} with superconducting qubits that gave a strong motivation for this 
study. 
We shall undertake a detailed
analysis of them with our methods. An analysis of their experiment from SDE perspectives have already been given in \cite{apoorva2017,apoorva2018}. Likewise, Konrad and collaborators have experimentally investigated quantum walks using orbital angular momentum of classical light
\cite{konradqwalk2013}. We plan to analyse their work from our perspectives. As already mentioned, the photon number distributions studied and
discussed at length in \cite{alter1,alter2,alterbook2001} need to be reformulated in our language. Even an analysis based on gaussian QND measurements, as done in sec.{\ref{sec:gaussianpovm}) should be valuable.

On the theoretical side, the physical meaning of the martingale convergence as already attempted in \cite{bauerlong2013} should be re-investigated in a more transparent way.

\acknowledgments
We acknowledge many useful and informative discussions with M.D. Srinivas, Rama Govindarajan, R. Vijayraghavan, Apoorva Patel, A.R. Usha Devi,
Sai Vinjanampathy  and members of quantum@iitm, in particular Arul Lakshminarayan, Vaibhav Madhok and Suresh Govindarajan. Both of us
gratefully acknowledge hospitality at TIFR-Hyderabad where major parts of this work were completed.


\begin{thebibliography}{}
\bibitem{born1926a} Max Born, Zeitschrift fur Physik, 37, p.863-7(1927)
\bibitem{born1926b} Max Born, Zeitschrift fur Physik, 38, p.803-27(1927)
\bibitem{paisborn1982} Abraham Pais, Science, 218, p.1193-8(1982).
\bibitem{heisenqm1925} W. Heisenberg, Zeit.f.Physik, 33 p.879(1925); English translation in \cite{vdWaerden}
\bibitem{bornjordanqm1925} M. Born and P. Jordan, Zeit.f.Physik 34 p.858-888. The full english translation by D.H. Delphenic. The version
included in \cite{vdWaerden} mysteriously omits the crucial chapter 4.
\bibitem{vdWaerden} B.L. van der Waerden, {\it Sources of Quantum Mechanics}, Dover Publications, New York.
\bibitem{einstein1917} A. Einstein, Physikalische Zeitschrift 18, p.121(1917)
\bibitem{jordanfoundqm1927} P. Jordan, Zeit.f.Physik 39,p.809-38(1927).
\bibitem{vNmathfound} J.von Neumann, {\it Mathematical Foundations of Quantum Mechanics}, Princeton University Press.
\bibitem{aduncan2025} Anthony Duncan, {\it Von Neumann's 1927 Trilogy on the Foundations of Quantum Mechanics},arXiv:2406.02149v2(27 May 2025).
\bibitem{JvN1927a} John von Neumann,{\it Mathematische Begr\"undung der Quantenmechank}, Konigliche Gesellschaft der Wissenschaften zu 
G\"ottingen.Mathematisch-physikalische Klasse.Nachrichten, p.1-57(1927); English translation 
{\it The mathematical foundation of Quantum Mechanics}\cite{aduncan2025}  
\bibitem{JvN1927b} John von Neumann, {\it Wahrscheinlichkeits-theoretischer Aufbau der Quantenmechank}, Konigliche Gesellschaft der 
Wissenschaften zu G\"ottingen.Mathematisch-physikalische Klasse.Nachrichten, p.245-272(1927); English translation 
{\it Probability-theoretic construction of Quantum Mechanics}\cite{aduncan2025}  
\bibitem{heisenunc1927} W. Heisenberg, Zeit.f.Physik, 43,p.172-98(1927);English translation in {\it Quantum Theory and Measurement}, Ed. J.A. Wheeler and W.H. Zurek, Princeton University Press,1983.
\bibitem{bohrqpost1928} N. Bohr, Nature 121, p.580-90(1928).
\bibitem{comptonsimons1925} A. Compton and D. Simons, Phys.Rev. 26(1925).
\bibitem{brakha1992} Vladimir B. Braginsky and Farid Ya. Khalili, {\it QUANTUM measurement}, Cambridge University Press, 1992.
\bibitem{rnsen2022} R.N. Sen, {\it von Neumann's book, the Compton-Simon experiment and the collapse hypothesis}, arXiv:2201.01299[physics.hist-ph].
\bibitem{luders1951} Gerhard L\"uders, Ann.Phys.(Leipzig) 8 p.322-328(1951); English translation in Ann.Phys.(Leipzig) 15, No 9,p.663(2006).
\bibitem{borneinsteinletts} {\it Born-Einstein Letters,1916-1955: Friendship, Politics, and Physics in uncertain times}, Macmillan Science
\bibitem{thornepreface} Diana Buchwald and Kip S.Thorne, {\it Preface to the New Edition of Born-Einstein Letters.}.
\bibitem{hartlesingle1968} J.B. Hartle, {\it Quantum Mechanics of Individual Systems}, Am.J.Phys. 36 p.704(1968). 
\bibitem{alterbook2001} O.Alter and Y.Yamomoto, {\it Quantum Measurements of a Single System},,Wiley Interscience, 2001.
\bibitem{alter1} O. Alter and Y. Yamamoto, Phys.Rev.Letts. 74(21), p.4106 (1995).
\bibitem{alter2} O. Alter and Y. Yamamoto, Fortschritte der Physik, 46(6-8), p.817(1998).
\bibitem{gisinprl1984} N.Gisin, Phys.Rev.Lett. 52 No 19 p.1657-1660(1984).
\bibitem{diosi1988} L. Diosi, J.Phys. A Math.Gen. p.2885(1988).
\bibitem{gisinpercival1992} N. Gisin and I.C. Percival, J.Phys. A Math.Gen. 25 p.5677(1992).
\bibitem{frohlichqm2023} J\"urg Fr\"ohlich, Zhou Gang, and Alessandro Pizzo, {\it A completion of quantum mechanics}, arXiv:2303.11112[math-phy]
\bibitem{frohlichjumps2025} J\"urg Fr\"ohlich, Zhou Gang, and Alessandro Pizzo, {\it A theory of quantum jumps}, Comm.Math.Phy. 8, p.195(2025).
\bibitem{jacobscontmeas2006} Kurt Jacobs and Daniel A Steck, {\it A straightforward introduction to continuous quantum measurement}, Contemporary Physics Vol 47 (2006).
\bibitem{weakrepeat2014} N.D. Hari Dass, {\it Repeated weak measurements on a single copy are invasive}, arXiv:1406.0270(quant-ph).
\bibitem{maassen2006} Hans Maassen and Burkhard K\"ummerer, {\it Purification of Quantum Trajectories}, IMS Lecture Notes on Dynamics and Statistics Vol 48(2016) 252-261.
\bibitem{nielsen2001} Michael Nielsen, {\it Characterising mixing and measurement in quantum mechanics}, Phys.Rev. A 63, 022114, 2001.
\bibitem{bauershort2011} Michel Bauer and Denis Bernard, {\it Convergence of repeated nondemolition measurements and wavefunction collapse}, 
Phys Rev A, 84 044103(2011).
\bibitem{bauerlong2013} Michel Bauer, Tristan Benoist, and Denis Bernard, {\it Repeated Quantum Non-Demolition Measurements: Convergence and 
Continuous Time Limit}, Ann.Henri Poincar\'e 14 p.639-679,2013.
\bibitem{amini2011} H. Amini, P. Rouchon, and M. Mirrahimi, {\it Design of Strict Control-Lyapunov Functions for Quantum Systems with QND Measurements}, arXiv:1103.1365v2.
\bibitem{stinespring1955} W.F. Stinespring, \it{Positive Functions on $C^*$-algebras.}, Proceedings of the American Mathematical Society, 6,
p.211-216(1955).
\bibitem{naimark1943} M.A. Naimark, Izv.Akad.Nauk SSSR, ser.Mat.7, 285(1943).
\bibitem{bhatia1997} R. Bhatia, {\it Matrix Analysis}, Graduate Texts in Mathematics, Vol.169, Springer, New York,1997.
\bibitem{bohmqt1989} David Bohm, {\it Quantum Theory}, Dover Publications, 1989.
\bibitem{paroanu2021} M. Cattaneo and G.S. Paraoanu, Advanced Quantum Technologies, Vol.4, No 11(2021).
\bibitem{arthurskelly1965} E. Arthurs and J.L. Kelly,Jr., Bell System Tech. J, 44(1965) 725.
\bibitem{quopfixed2002} A. Arias, A. Cheondas, and S. Gudder, J.Math.Phys 43, 5872(2002).
\bibitem{belavkinmultivar1992} V.P. Belavkin, J. Multivar. Anal,42, p.171-201,(1992).
\bibitem{belavkincmp1992} V.P. Belavkin, Commun.Math.Phys. 146,611-635,(1992).
\bibitem{ushaparth2017} K.R. Parthasarathy and A.R. Usha Devi, {\it Asymptotic Spectral Stability of the Gisin-Percival State Diffusion}, arXiv:1707.08157v2[quant-ph].
\bibitem{adler2001} S.L. Adler et al, J.Phys. A34 p.8795(2001).
\bibitem{stockton2004} R. van Handel, J.K. Stockton, and H.Mabuchi, Phys.Rev. A 70, p.022106(2004).
\bibitem{stockton2005} J.K. Stockton, R. van Handel, and H. Mabuchi, IEEE T. Automat.Contr. 50 ,p.768(2005).
\bibitem{qzeno1977} B. Misra and E.C.G. Sudarshan, {\it Zeno's paradox in quantum theory}, J.Math.Phys. 18(4) p.756(1977).
\bibitem{invqzeno} Y. Aharonov and M. Vardi, Phys.Rev. D21(8) p.2235(1980).
\bibitem{vijayexpt2016} K.W. Murch, R. Vijay, and I. Siddiqi, {\it Weak measurement and feedback in superconducting quantum circuits}, in
{\it Superconducting devices in Quantum Optics}, R. Hadfield and G. Johansson(Eds),p.163, Springer,2016.
\bibitem{apoorva2017} A. Patel and P. Kumar, {\it Weak measurements, quantum state collapse and the Born rule}, Phys.Rev. A96 022108(2017).
\bibitem{apoorva2018} Parveen Kumar et al, {\it Quantum Trajectory Distribution for Weak Measurement of a superconducting qubit: Experiment
meets Theory}, arXiv:1804.03413v1[quant-ph].
\bibitem{konradqwalk2013} Sandeep K. Goyal et al {\it Implementing Quantum Walks Using Orbital Angular Momentum of classical light}, Phys.Rev.Lett. 110 263602(2013).
\end{thebibliography}
\end{document}